\documentclass[journal,final,letter,onecolumn,12pt]{IEEEtran}

\usepackage{graphicx}
\usepackage{amsmath,bbm}
\usepackage{epsfig}

\def\bz{{\bf z}}
\def\CC{{\cal C}}

\def\CK{{\cal K}}
\def\bu{{\bf u}}
\def\bv{{\bf v}}
\def\bw{{\bf w}}
\def\ie{{\em i.e., }}
\def\eg{{\em e.g., }}
\def\A{{\cal A}}
\def\TP{\tilde{P}}

\def\tw{\tilde{g}}
\def\TW{\tilde{G}}

\newtheorem{theorem}{Theorem}
\newtheorem{corollary}{Corollary}
\newtheorem{lemma}{Lemma}

\begin{document}
\title{Information Propagation Speed in Mobile and Delay Tolerant Networks
\thanks{Part of this work will be presented in ``Information Propagation Speed in Mobile and Delay Tolerant
  Networks'', P. Jacquet, B. Mans and G. Rodolakis, IEEE Infocom, Rio de Janeiro, Brazil, April, 2009.}
}

\author{
Philippe Jacquet, Bernard Mans and Georgios Rodolakis
\thanks{
P. Jacquet is with INRIA, 78153 Le Chesnay, France. E-mail: philippe.jacquet@inria.fr~.}
\thanks{
B. Mans and G. Rodolakis are with Macquarie University, 2109 NSW, Australia.
E-mails: bernard.mans@mq.edu.au, \mbox{georgios.rodolakis@mq.edu.au~.}}}

\markboth{Submitted to IEEE Transactions on Information Theory}{Submitted to
IEEE Transactions on Information Theory}

\maketitle

\begin{abstract}
The goal of this paper is to increase our understanding of the fundamental performance limits of mobile and Delay Tolerant Networks (DTNs), where end-to-end multi-hop paths may not exist and communication routes may only be available through time and mobility. We use analytical tools to derive generic theoretical upper bounds for the information propagation speed in large scale mobile and intermittently connected networks. In other words, we upper-bound the optimal performance, in terms of delay, that can be achieved using any routing algorithm. We then show how our analysis can be applied to specific mobility and graph models to obtain specific analytical estimates. In particular, in two-dimensional networks, when nodes move at a maximum speed $v$ and their density $\nu$ is small (the network is sparse and surely disconnected), we prove that the information propagation speed is upper bounded by ($1+O(\nu^2))v$ in the random way-point model, while it is upper bounded by $O(\sqrt{\nu v} v)$ for other mobility models (random walk, Brownian motion). We also present simulations that confirm the validity of the bounds in these scenarios.
Finally, we generalize our results to one-dimensional and three-dimensional networks.
\end{abstract}

\baselineskip 19 pt

\section{Introduction}
\label{Sect:intro}

Recent research has highlighted the necessity and the significance of mobile ad hoc networks where end-to-end multi-hop
paths may not exist and communication routes may only be available
through time and mobility. Depending on the context, these networks
are commonly referred as Intermittently Connected Networks
(ICNs) or Delay Tolerant Networks (DTNs).

While there is a large body of work on understanding the fundamental properties and performance limits of wireless networks under the assumption that connectivity must be maintained (\emph{e.g.}, since the seminal work by Gupta and Kumar~\cite{GK00}), there are only few results on the properties of intermittently connected or delay tolerant networks (\emph{e.g.},~\cite{soda,KY08,KY08infocom,PMCC07}). Most of the effort has been
dedicated to the design of efficient routing protocols (see~\cite{Z06} for a survey)
and comparative simulations, using specific mobility models or concrete traces
(\eg \cite{SPG06,ZKLTZ07}). A complete understanding of what one can expect for optimal
performance (\eg through theoretical bounds) is still missing for
many realistic
models.

In this context, the objective of the paper is to evaluate the maximum speed at which a piece of
information can propagate in a mobile wireless network.
A piece of information is a packet (of small size) which
can be transmitted almost instantaneously between two nodes in range.
If the network
is connected ({\it i.e.}, an end-to-end multi-hop path exists) information
moves at a rather high speed, which can be considered infinite compared
to the mobility of
the nodes.

We consider a network made of $n$ nodes moving in a domain of size $\A$ (in two dimensions a square area), under the unit
disk graph model ({\it i.e.}, nodes are neighbors when their distance
is smaller than one). In order to study the properties of DTNs that are relevant to the field of applications,
we are interested in very sparse networks and we are investigating
the case where the node density
$\frac{n}{\A}$ is small. 
Indeed, most applications for DTNs
are required to work for sparse mobile ad hoc networks (\emph{e.g.},~\cite{SPG06,Z06,ZKLTZ07}), where
intermittent connectivity is due to node mobility and to
limited radio coverage.
In these cases, the mobile network is almost always
disconnected, making information propagation stall as
long as the node mobility does not allow the information to jump to
another connected component. The information is either
transmitted or carried by a node (requiring a {\em
store-carry-and-forward} routing model).  Thus, a ``path'' is an
alternation of packet transmissions and carriages, that connects a
source to a destination, and is better referred (from now on) as a
{\bf journey}.
Informally, our aim is to find the shortest journey (in time) that
connects any source to any destination in the network domain, in order to derive
the overall propagation speed.

In terms of related work on the information propagation speed in wireless networks, the problem has been studied in static networks. Zheng~\cite{zheng} showed that there is a constant upper bound on the information diffusion rate in large wireless networks. Recently,
Xu and Wang~\cite{XW08} proved that there is a unified upper bound on the maximum propagation speed in large wireless networks, using unicast or broadcast. The article~\cite{MASS} evaluates analytical upper bounds on the packet propagation speed using opportunistic routing. In contrast, our main focus here is to evaluate the information propagation speed in mobile and intermittently connected networks.

Taking into account the node mobility, some recent papers have presented initial results on the theoretical properties of intermittently connected networks, \emph{e.g.},~\cite{soda,GNK05,KY08,KY08infocom,PMCC07,ZNKT07}.
The papers~\cite{GNK05,ZNKT07} analyze the delay of common routing schemes, such as epidemic routing, under the assumption that the inter-meeting time between pairs of nodes follows an exponential distribution. The authors of~\cite{PMCC07} took a graph-theoretical approach in order to upper bound the time it takes for disconnected mobile networks to become connected through the mobility of the nodes. This work uses an Erd\"os-R\'enyi network model, where
the node connections are done independently of the actual topology of the network.
In this paper, we will depart from this model in order to
integrate the topological nature of the network, for an instance of $n$ mobile
nodes, first in a square map of size $\A$
connected according to the unit disk graph model, and then generalized to a map of dimension $D$.

In~\cite{soda}, an interesting model of dynamic random geometric graphs (based on a random walk mobility model) leads to the first precise asymptotic results on the connectivity and disconnectivity periods of the network. Unfortunately, this methodology cannot be extended to evaluate the fastest possible information propagation.
In~\cite{KY08,KY08infocom}, Kong and Yeh studied the information dissemination latency in large wireless and mobile networks, in constrained i.i.d. mobility and Brownian motion models. They showed that, when the network is not percolated (under a critical node density threshold), the latency scales linearly with the Euclidean distance between the sender and the receiver, while the latency scales sub-linearly in the super-critical case where the network is percolated. A question that remains to be answered is to find precise estimates on the constant upper bounds of the information propagation speed in intermittently connected mobile networks.
In~\cite{JM07}, the authors present an initial analytical upper bound on the achievable information propagation speed in an infinite network model. Here, we present the first analytical results in the more realistic (and significantly more difficult) case of a large scale but finite mobile network model, in order to prove rigorous upper bounds on the maximum achievable information propagation speed. Moreover, we derive our theoretical bounds on a more general mobility model than those used in the literature, while we also compare our analytical results with simulations.

More precisely, our main contributions are the following:
\begin{itemize}
\item we present a new probabilistic model of space-time journeys of packets of information in delay tolerant networks;
\item we upper bound the optimal performance that can be achieved using any routing algorithm in finite two-dimensional mobile networks and we derive theoretical bounds on the information propagation speed, depending on the node density and the network mobility;
\item we generalize our results for bounded multi-dimensional networks;
\item we verify the accuracy of our bounds via simulations.
\end{itemize}

The rest of the paper is organized as follows. We first analyze in detail the case of two-dimensional networks; in Section~\ref{Sect:model}, we introduce the network and mobility model, we define the information propagation speed metric, and we discuss our main results and the methodology.
In Section~\ref{Sect:analysis}, we present the detailed analysis and the proof of our theoretical upper bounds. We derive asymptotic estimates for the propagation speed in sparse networks in Section~\ref{sect:sparse}. We then generalize our results in a more general model of multi-dimensional networks in Section~\ref{Sect:multi}.
We illustrate the behavior of the bounds depending on the network and mobility parameters (such as the node density and change of direction rate) in Section~\ref{sect:slowness}.
We compare the analytical bounds with simulation measurements in Section~\ref{Sect:sim}. We conclude and propose some directions for further research in Section~\ref{Sect:conclusion}.

\section{Model and Overview of Main Results in Two-Dimensional Networks}
\label{Sect:model}
\subsection{Mobile Network Model}

In the two-dimensional case, we consider a network of $n$ nodes in a square area of size
$\A=L\times L$. The nodes are enumerated from 1 to $n$.
In the next section, we will analyze the case where both $n,L\to \infty$ such that the node density $\nu=\frac{n}{\A}$ tends to a (small) constant.

Initially, the nodes are distributed uniformly at random.
Every node follows an i.i.d. random trajectory, reflected on the
borders of the square like billiard balls.
The nodes change direction at Poisson rate~$\tau$ and keep a
uniform speed between direction changes. The motion direction angles are
uniformly distributed between $0$ and $2\pi$.
When $\tau > 0$, we have a random walk model; when $\tau\to\infty$ we are on the Brownian limit; when $\tau\to 0$ we
are on a random way-point-like model, since nodes travel a distance of order $L$ before changing direction.

The billiard model is equivalent to considering an infinite area
made of mirror images of the original square: a mobile node moves
in the original square while its mirror images move in the mirror
squares.
The fact that a node bounces on a border is strictly equivalent to
crossing it without bouncing, while its mirror
image enters the square.
With this perspective, the trajectory of a node is equivalent to
a free random trajectory in the set of mirror images of the original
square, while the nodes remain distributed uniformly at random.

We adopt the unit-disk model: two nodes at a distance
smaller than one can exchange information.
The average number of
neighbors per node is therefore smaller (or equal) than $\pi\frac{n}{\A}$.
In~\cite{XK04}, Xue and Kumar have shown that if the average number of
neighbors is smaller than $0.074 \log n$,
then the network is almost surely disconnected when $n$ is large.
In order to study the properties of delay tolerant networks in the context of their applications, we need to look at
sparse networks. Therefore, we assume that the number of nodes $n$ tends to infinity at the same rate as the area of the network domain square (so that the node density remains constant), and we investigate
the case where the node density
$\frac{n}{\A}$ is small.

Since we are interested in computing upper bounds on the best possible information propagation, we do not consider here the effects of buffering or congestion. Indeed, we assume that a piece of information, \emph{i.e.}, a packet of small size
can be transmitted instantaneously between two nodes in range. Even under these assumptions, we are able to derive finite bounds on the information propagation speed. We note that these assumptions do not affect the validity of our upper bounds, since they correspond to an ideal scenario with that respect; this allows us to capture the fundamental performance limit of DTNs based solely on the network mobility and topology. Moreover, in the case of very sparse mobile networks, the previous assumptions do not impact on the accuracy of our results, since information transmission occurs much faster than the speed of the mobile nodes.

\subsection{Information Propagation Speed and Main Results}
Our main result is the evaluation of a generic upper bound of the information propagation speed (presented later in Theorem~\ref{Theo:upper} in this section), which in turn allows us to obtain specific bounds for particular models.

In order to evaluate the fastest
possible information propagation, we establish a probabilistic space-time model of journeys of packets of information in delay tolerant networks that contains
all possible ``shortest'' journeys: the full epidemic broadcast. We call the
information, the {\it beacon}. Every time a new node is in range of a
node which carries a copy of the beacon, the latter node transmits
another copy of the beacon to the new node. In our model, journeys are expressed as space-time trajectories, since store-carry-forward routing also implies that we must take into account the time dimension.

To prove our main theorem in Section~\ref{Sect:analysis}, we decompose the packet journeys into independent segments and we evaluate the Laplace
transform of the journey probability density. From the Laplace transform, we are able to establish an upper bound on the
average number of journeys arriving to a point~$\bz$ before a time~$t$, where~$\bz$ is a 2D space vector expressing the spatial distance from the source that emitted the beacon. More precisely, we are interested to find when the density of journeys becomes~0 almost surely. We notice that a zero probability of reaching a given point in space in a given amount of time implies an upper bound on the information propagation speed. In order to evaluate a constant bound, we will consider the asymptotic case where the distance from the source and the time both tend to infinity. Hence, using our approach, we obtain theoretical bounds on the information propagation speed by computing the smallest ratio of the distance over the given time, which yields a journey probability of zero.
The asymptotic approach must be interpreted in the following sense: we evaluate the information propagation speed to a distance which is a large multiple of the maximum radio range. In fact, in Section~\ref{Sect:sim}, we will see that the propagation speed quickly converges to a constant value as soon as the distance between the source and the destination is simply larger than the radio range.
Similarly, in~\cite{KY08}, in the case of disconnected mobile networks, the authors show that the information propagation latency scales linearly with the distance in the same asymptotic setting.

Therefore, the concept of propagation speed is probabilistic.
To express the previous discussion using mathematical notations, let us consider that the beacon starts at time $t=0$ on a node at coordinate
$\bz_0=(x_0,y_0)$. Let us initially consider (for simplicity) a destination node
that stays at coordinate $\bz_1=(x_1,y_1)$. Let
$q_\nu(\bz_0,\bz_1,t)$ denote the probability that the destination receives
the beacon before time~$t$.
A scalar~$s_0>0$ is an upper bound for the propagation speed, if for all~$s>s_0$,
$\lim q_\nu\left(\bz_0,\bz_1,\frac{|\bz_1-\bz_0|}{s}\right)=0$ when $|\bz_1-\bz_0|\to\infty$, with $|.|$ denoting the Euclidean norm.
For example, if we prove that $q_\nu(\bz_0,\bz_1,t)<\exp(-a|\bz_1-\bz_0|+bt+c)$, then quantity~$\frac{b}{a}$ is a propagation speed
upper bound.

Using the previously described methodology, we will prove the following main theorem, which expresses our generic upper bound on the information propagation speed in terms of different values of the network and mobility parameters.

\begin{theorem}
  For a network in a square area $\A=L\times L$,
  where the number of nodes $n\to\infty$ and $L\to\infty$ such that the node density $\nu=\frac{n}{\A}$ remains constant, an upper bound on the information propagation speed is the smallest ratio:
$$
\min_{\rho,\theta>0}\left\{\frac{\theta}{\rho}~\text{with}~
\theta=\sqrt{\rho^2 v^2+\left(\tau+\frac{\frac{n}{\A}4\pi v I_0(\rho)}{1-\frac{n}{\A}\pi\frac{2}{\rho}I_1(\rho)}\right)^2}-\tau\right\},
$$
where $v$ is the maximum node speed, $\tau$ is the node direction change rate, while $I_0()$ and $I_1()$ are {\em modified Bessel functions} (see~\cite{AS65}), defined respectively by:
$I_0(x)=\sum_{k\ge 0}(\frac{x}{2})^{2k}\frac{1}{(k!)^2}$,
and
$I_1(x)= \sum_{k\ge 0}(\frac{x}{2})^{2k+1}\frac{1}{(k+1)!k!}.$
\label{Theo:upper}
\end{theorem}
\paragraph*{Remark} As we will see, quantities $\rho$ and $\theta$ correspond to the parameters of the Laplace transform of the journey probabilities. Quantity $\rho$ is expressed as an inverse of distance and quantity $\theta$
is expressed as an inverse of time, therefore the ratio $\frac{\theta}{\rho}$ has the
dimension of a speed.

Since quantities $I_0(x)$ and $\frac{2}{x}I_1(x)$ are both greater than 1,
the previous expression has meaning
when $\frac{n}{\A}<\frac{1}{\pi}$.
Above this density threshold,
the upper bound for the information propagation speed is infinite.
Such a behavior is expected, since it is known that there exists a critical density above which the graph is fully connected or at least
percolates (\ie there exists a unique infinite connected component with non-zero probability)~\cite{MR96}.
The infinite component implies an infinite information propagation speed according to our definition.
The exact value of the critical density is unknown, although there are known bounds and numerical estimates~\cite{DTH02}.
However, in the context
of mobile delay tolerant networks, we are interested to
analyze the sub-critical case.
We note that the critical threshold obtained from our analysis is smaller than the critical percolation density.

Theorem~1 gives a concise upper bound on the information propagation speed, which we will illustrate in detail in Section~\ref{sect:slowness}. In order to give a more intuitive understanding of the fundamental performance limits of the information propagation speed, we derive the following corollaries expressing the qualitative behavior of the upper bound when the node density tends to $0$. This case models very sparse mobile wireless networks, which as discussed are of special interest in the context of delay tolerant networks.
\begin{corollary}
When nodes move at speed $v>0$ in a random walk model
(with node direction change rate $\tau >0$), and when the square length
$L\to\infty$, but such that the node density $\frac{n}{\A}\to 0$, the propagation speed
upper bound is asymptotically equivalent to
$O(\sqrt{\frac{n v}{\A\tau}} v)$.
\end{corollary}

It is important to notice that the speed diminishes with the square
root of the density $\nu$.

A special case corresponds to $\tau=0$, which is a pure billiard model (nodes
change direction only when they hit the border).
\begin{corollary}
When nodes move at speed $v>0$ with $\tau=0$, and when
$L\to\infty$, but with node density $\nu=\frac{n}{\A}\to 0$,
the propagation speed upper bound is
$(1+ O(\nu^2))v$.
\end{corollary}

It turns out that the propagation speed upper bound
at the limit is $v$. This is rather surprising because we would expect that
the propagation speed would tend to zero when $\nu\to 0$.

We note that the above results do not contradict the results
of~\cite{PMCC07}, although they can not be directly compared, since
a unit disk graph cannot be modeled like an Erd\"os-R\'enyi graph. Indeed
if nodes $A$ and $B$ are connected to a same third node $C$, then both will be
connected with a much higher probability than the probability we would
had if they were in an Erd\"os-R\'enyi graph. On the other hand, our analysis in fact confirms the results of~\cite{KY08}, which imply that the information propagation speed tends to a constant and finite value in intermittently connected networks; our results give the first estimates of this finite information propagation speed.

\section{Analysis (Proof of Theorem~1)}
\label{Sect:analysis}

\subsection{Methodology and Journey Analysis}

Our analysis is based on a segmentation of journeys between the source and
the destination. Formally, a journey is a space-time trajectory of the beacon
between the source and the destination. In the following,
we first decompose journeys into segments (\ie space-time vectors) which model the node trajectories and the beacon transmissions in Section~\ref{sect:segment}. Our aim is to decompose journeys into independent segments, therefore a technical difficulty comes from the dependence in the node emissions and movements (for instance, the direction of an emission depends on the direction of the node movement). However, we see how we can use an independent segment decomposition in Section~\ref{sect:dependent}, in order to upper bound the journey probabilities.
We then calculate the Laplace transforms of each individual segment, and, making use of the journey decomposition, we deduce the Laplace transform of the probability density of each journey in Section~\ref{sect:laplace} for a fixed length sequence of segments. Finally, an asymptotic analysis on the journey Laplace transform (for large scale networks), based on Poisson generating functions, allows us to compute when the journey probability density tends to zero, and consequently evaluate an upper bound on the information propagation speed in Section~\ref{sect:speed}.

We assume that time zero is when
the source transmits, and we will check at what time $t$ the beacon
is emitted at distance smaller than one to the destination at
coordinate $\bz=(x,y)$.
The beacon can take many journeys in parallel,
due to the broadcast nature of radio transmissions, and the fact that
the beacon stays in the memory of each emitter (and therefore can be
emitted several times in the trajectory of a mobile node).  In a first
approach and in order to simplify, we assume that the destination
is fixed; however, we will later see that the destination motion does not affect our results.

We will only consider simple journeys, \ie journeys which never return
twice through the same node. This restriction
does not affect the analysis, since if a journey arrives to the destination
at time~$t$, then we can extract a simple journey from this journey which
arrives at time~$t$ too.

Let $\CC$ be a simple journey. Let $Z(\CC)$ be the terminal point.
Let $T(\CC)$ be the time at which the journey terminates. Let $P(\CC)$
be the probability of the journey $\CC$.
In the following, we consider a journey as a discrete
event in a continuous set of all possible journeys in space-time, and we convert the probability weight $P(\CC)$ to a probability density.

Assuming that there are $n$ nodes in the network,
we call $p_n(\bz_0,\bz_1,t)$ the density of journeys starting from $\bz_0$
at time~0, and arriving at~$\bz_1$ before time~$t$:
$$
p_n(\bz_0,\bz_1,t)=\lim_{r\to 0}\frac{1}{\pi r^2}\sum_{|\bz_1 -Z(\CC)|<r,T(\CC)<t}P(\CC)~.
$$

\subsection{Journey Segmentation}\label{sect:segment}
Let us consider a journey where the beacon is carried by $k+1$ nodes
$\ell_0,\ell_1,\ell_2,\ldots,\ell_k$. The node $\ell_0$ is the source.
Let $\ell_1$ be the first node that receives the beacon from the source,
$\ell_2$ the node
that receives the beacon from $\ell_1$, {\it etc}.
We call the $k$-tuple $(\ell_1,\ell_2,\ldots,\ell_k)$
the journey relay sequence $N(\CC)$.
\begin{lemma}
The probability distribution of the journey $\CC$ only depends on the cardinality $|N(\CC)|$.
\label{lem:card}
\end{lemma}
\begin{proof}
Since node motions are i.i.d., any node in $N(\CC)$ can be interchanged
with any other node.
\end{proof}
Consequently, we can split the journey $\CC$ into segments
$(s_0,s_1,s_2,\ldots,s_k)$, where the segments $s_i$ are random space-time vectors, and where $s_i$
is the space-time vector that starts with the event:
``the beacon is received by $\ell_i$''. In the special case of $i=0$, the event is the origin
of the journey.

To compute the probability distribution of the segments, we notice that $s_i$ corresponds to a space-time motion trajectory of mobile node $\ell_i$ (the trajectory can be possibly zero if the node immediately retransmits the beacon),
and a space-time vector of the beacon transmission via radio (where the time component of a transmission is zero).
Therefore, in order to decompose a journey, we define two kinds of segments modeling the described situations:
\begin{itemize}
\item \emph{emission segments} $S_e(\bu,\bv)$: the node
transmits immediately after receiving the beacon;
$\bv$ is the speed of the node that just received the beacon, and $\bu$ is the
emission space vector and is such that $|\bu|\le 1$;
\item \emph{move-and-emit segments} $S_m(\bu,\bv,\bw)=M(\bv,\bw)+\bu$: $M(\bv,\bw)$ is the
space-time vector corresponding to the motion of the node carrying the beacon,
where $\bv$ is the initial
vector speed of the node when it receives the beacon and
$\bw$ is the final speed of the node just
before transmitting the beacon; the vector $\bu$ is the emission space vector
which ends the segment.
\end{itemize}

With the following lemma, we prove that the vector $\bu$ which ends the move-and-emit segments can be restricted to unitary segments.
\begin{lemma}
In a ``fastest" journey decomposition (\ie with respect to an upper bound on the information propagation speed),
move-end-emit segments $S(\bu,\bv,\bw)$ can be restricted to unitary emission
vectors: $|\bu|=1$.
\label{lem:fastest}
\end{lemma}
\begin{proof}
First, assume that $\ell_{i}$ and $\ell_{i+1}$ are not neighbors when $\ell_i$
receives the beacon. The earliest time at which $\ell_{i+1}$ will
receive the beacon from $\ell_i$ is when both become neighbors, \ie
when their distance is just equal to~1; therefore, the emit vector
is unitary. Conversely, if $\ell_{i}$ and $\ell_{i+1}$ are already
neighbors when $\ell_i$ receives the beacon, 
then $\ell_{i+1}$ can receive the packet immediately after $\ell_i$ and
the segment would be an emission segment instead.
\end{proof}

Since we want to check when a beacon can be emitted at distance less than one from the destination, we do not include the last emission in our journey definition; therefore, the last segment $s_k$ corresponds only to the space-time motion trajectory of node $\ell_k$ (or simply, a motion equal to zero).

\subsection{Decomposition into Independent Segments}\label{sect:dependent}
In this section, our aim is to decompose journeys into independent \emph{emission} and \emph{move-and-emit} segments.
However, there is a dependence in successive node emissions and movements; for example, a node moving faster meets more nodes than a slower mobile node; similarly, the probability of a meeting between two nodes is in fact proportional to the relative speed between the nodes, hence two nodes that meet are more likely to move in (almost) opposite directions; therefore, the direction of an emission depends on the direction of the node movement.
To overcome these difficulties, we will in fact work with an upper bound on the journey probability densities, and we show that this upper bound can be decomposed into independent segments.

Thus, our objective is to compute an upper bound on $P(\CC)$, the probability density
that
a journey $\CC$ exists.
For a fixed journey relay sequence $N(\CC)$ of size $k$,
the probability density is a vector in ${R}^{3k}$. Based on the journey decomposition, we have the expression
$P(\CC)=p(s_0|s_1)p(s_1|s_2)\cdots p(s_{k-1}|s_k)p(s_k)$, where
$p(s_{i}|s_{i+1})$ is the conditional probability density of segment $s_i$, given the next segment $s_{i+1}$.
We have the following expressions for the conditional probabilities, for all possible combinations of emission and move-and-emit segments:
\begin{itemize}
\item $p(S_e(\bu_1,\bv_1)|S_e(\bu_2,\bv_2))=P(\bu_1)P(\bv_1)\frac{1}{\A}$;
\end{itemize}
this is the probability of emission segment $S_e(\bu_1,\bv_1)$, when we know the next segment (here, an emission segment):
    $P(\bu_1)$ is the probability density of $\bu_1$ inside the unit disk (emissions are equiprobable in the unit disk, hence $P(\bu_1)=\frac{\partial}{\partial |\bu_1|}\frac{\pi |\bu_1|^2}{\pi}=2|\bu_1|$), $P(\bv_1)$ is the probability that the node moves at speed~$\bv_1$, and~$\frac{1}{\A}$ is the density of presence of a node on the second segment (to make the emission possible); there is no dependence on the parameters $\bu_2,\bv_2$ of the second segment, since the node receiving the packet re-emits it immediately to one of its neighbors (there is no new meeting);
\begin{itemize}
\item $p(S_e(\bu_1,\bv_1)|S_m(\bu_2,\bv_2,\bw_2))=P(\bu_1)P(\bv_1)\frac{1}{\A}$, for the same
reason;
\item $p(S_m(\bu_1,\bv_1,\bw_1)|S_e(\bu_2,\bv_2))= P(M(\bv_1,\bw_1))P(\bu_1)P(\bu_2)P(\bv_2)\max\{0,\bu_1\cdot(\bw_1-\bv_2)\}\frac{1}{\A}$;
\end{itemize}
    this is the probability of the move-and-emit segment $S_m(\bu_1,\bv_1,\bw_1)$, when we know the next segment (here, an emission segment);
    quantity $\max\{0,\bu_1\cdot (\bw_1-\bv_2)\}\frac{1}{\A}$ is the
average rate at which a node
carrying the beacon on the second segment enters
the neighborhood range of the previous node on the radius
$\bu_1$ with relative speed $\bv_2-\bw_1$ (see Appendix~\ref{ap:rate});
quantities $P(M(\bv_1,\bw_1)),P(\bu_1),P(\bu_2)$ and $P(\bv_2)$ correspond to the probabilities of the respective space and speed vectors (we note the dependence on the parameters $\bu_2,\bv_2$ expressing the probability of the second segment, since the first segment includes a node motion, during which the packet is carried before being transmitted to a new neighbor);
\begin{itemize}
\item $p(S_m(\bu_1,\bv_1,\bw_1)|S_m(\bu_2,\bv_2,\bw_2))=
P(M(\bv_1,\bw_1))P(\bu_1)P(\bu_2)P(M(\bv_2,\bw_2))\max\{0,\bu_1\cdot (\bw_1-\bv_2)\}\frac{1}{\A}$, for the same reason.
\end{itemize}

From the above, we notice that a journey cannot be directly decomposed into independent segments, because of the conditional probabilities. However, recall that in order to derive an upper bound on the information propagation speed, it suffices to compute when the probability of a journey becomes zero. Therefore, we can instead use an upper bound on the journey probabilities, and check when this upper bound becomes zero.
Based on the previous expressions, we can upper bound the conditional probabilities. Hence, an upper bound of the density of $\CC$ is $\TP(\CC)$, with
$\TP(\CC)=\TP(s_1)\TP(s_2)\cdots\TP(s_{k-1})\TP(s_k)$, and:
\begin{itemize}
\item $\TP(S_e(\bu,\bv))=P(\bu)P(\bv)\frac{1}{\A}$,
\item $\TP(S_m(\bu,\bv,\bw))=P(\bu)P(M(\bv,\bw))2\max\{v\}\frac{1}{\A}$,
\end{itemize}
where $\max\{v\}$ denotes the maximum node speed.

Looking at all the above equations, we observe that $\TP(s_i)\ge P(s_i|s_{i+1})$ for all $i$ and any combination of segments. Using the new segments probabilities, we have an upper bound journey model that can be decomposed into independent emission and move-and-emit segments.

\subsection{Journey Laplace Transform}\label{sect:laplace}

Let $\sigma=(\zeta,\theta)$ be a is an inverse space-time vector: $\zeta$ is a space vector with components expressed in inverse distance units, and $\theta$ is a scalar in inverse time units.
We define $\tw_k(\sigma)$ as the
Laplace transform of the upper bound density of a journey $\CC$ given that $N(\CC)$ is fixed of size $k$. In other words, we have by the Laplace transform definition:
$\tw_k(\sigma)=E(\exp(-\sigma \cdot (s_0+s_1+\cdots+s_k)))$, under the probability
weight $\TP$. Notice that the exponent
is the dot product of two vectors, and that this product is a pure scalar without dimension, since $(\zeta,\theta)$ is an inverse space-time vector.

\begin{lemma}
The Laplace transform of the upper bound journey density given that the relay
sequence $N(\CC)$ is fixed and is of length $k$, satisfies:
\begin{eqnarray*}
\tw_k(\sigma)&=&\left(2\max\{v\}E(e^{-\sigma \cdot (M(\bv,\bw)+\bu)})+E(e^{-\zeta\cdot \bu})\right)^k
\left(\frac{1}{\A}\right)^k \\
&&\times E(e^{-\sigma\cdot M(\bv,\bw)}),
\end{eqnarray*}
where $\max\{v\}$ denotes the maximum node speed.
\label{lem:lapl}
\end{lemma}
\begin{proof}
This is a direct consequence of the independence of segments in the
upper bound journey density model, which implies that the journey Laplace transform can be expressed as a product of the individual segment Laplace transforms. The first line expresses the Laplace transform of a sequence of $k$ emission \emph{or} move-and-emit segments, while the last term corresponds to the last segment
which corresponds only to a space-time motion trajectory.
\end{proof}

Let $\TW_n(\sigma)$ be the Laplace transform of the upper bound
density of all journeys in a network of size $n$ in a square map of
area size $\A$.
Now,
the remaining difficulty comes from the fact that $N(\CC)$ is not known or fixed.
To tackle this problem, we define the Poisson generating function:
$$
\TW(Z,\sigma)=\sum_{n\ge 0}\TW_n(\sigma)\frac{Z^n}{n!}e^{-Z}.
$$
\begin{lemma}
The following identity holds:
$$
\TW(Z,\sigma)=\sum_{k\ge 0}\tw_k(\sigma)Z^k~.
$$
\label{lm-poi}\end{lemma}
\begin{proof}
This is a formal identity. Quantity $\tw_k(\sigma)$ depends only
on the actual length of the relay sequence and not on the nodes
that are actually in the relay sequence (from Lemma~\ref{lem:card}), thus the Laplace transform of
the journeys that are made of $k$ segments is
$\frac{n!}{(n-k)!} \tw_k(\sigma)$, since $\frac{n!}{(n-k)!}$ is the
number of distinct relay sequences of size $k$.

This means that
$
[Z^n]~\TW(Z,\sigma)=\sum_{k\ge n}\frac{n!}{(n-k)!}\tw_k(\sigma)~,
$
and
$$
\begin{array}{rcl}
\TW(Z,\sigma)e^Z&=&\sum_{n\ge 0,k\ge n}\frac{n!}{(n-k)!}\tw_k(\sigma)\frac{Z^n}{n!}\\
&=&\sum_{k\ge 0} \tw_k(\sigma)e^Z.
\end{array}
$$
\end{proof}
\begin{corollary}
\label{cor:laplace}
We have (from Lemmas~\ref{lem:lapl} and~\ref{lm-poi}):
$$
\TW(Z,\sigma)= \frac{E(e^{-\sigma\cdot M(\bv,\bw)})}
{1-\left(2\max\{v\}E(e^{-\sigma \cdot(M(\bv,\bw)+\bu)})+E(e^{-\zeta\cdot \bu})\right)\frac{Z}{\A}}.
$$
\end{corollary}

Therefore, we can now evaluate the Poisson generating function $\TW(Z,\sigma)$ of the journey Laplace transform by combining the segments Laplace transforms. In the following lemma, we evaluate the expressions for the Laplace transforms.
\begin{lemma}
We have:
\begin{itemize}
\item $E(e^{-\zeta \cdot \bu})=2 \pi \nu I_0(|\zeta|)$ when
$\bu$ is unitary and uniform on the unit circle, with density $\nu$;
\item $E(e^{-\zeta \cdot \bu})=\nu \frac{2\pi}{|\zeta|}I_1(|\zeta|)$ when $\bu$ is uniform
in the unit disk, with density $\nu$;
\item When all speeds are of modulus equal to $v$, we have
$E(e^{-\sigma \cdot M(\bv,\bw)})=\frac{1}{\sqrt{(\theta+\tau)^2-|\zeta|^2v^2}-\tau}$;
\end{itemize}
where $I_1()$ and $I_0()$ are
modified Bessel functions.
\label{Lem:laplace}
\end{lemma}
\begin{proof}
See Appendix~\ref{ap:laplace}.
\end{proof}

\subsection{Information Propagation Speed Analysis}\label{sect:speed}

Our aim is to obtain an estimate of $p_n(\bz_0,\bz_1,t)$, \ie the upper bound
on the density of journeys that start at~$\bz_0$ at time~$0$ and end at~$\bz_1$
at time~$t$.
Let $p(Z,\bz_0,\bz_1,t)$ be the Poisson generating function of
$p_n(\bz_0,\bz_1,t)$, that is:
$p(Z,\bz_0,\bz_1,t,z)=\sum_n p_n(\bz_0,\bz_1,t)\frac{Z^n}{n!}e^{-Z}$.

\begin{lemma}
The generating function $p(Z,\bz_0,\bz_1,t)$ has positive coefficients.
\label{lem:positive}
\end{lemma}
\begin{proof}
From Lemma~\ref{lm-poi}:
$$p(Z,\bz_0,\bz_1,t)=\sum_{k \ge 0} p_k(\bz_0,\bz_1,t)Z^k.$$
\end{proof}

Hence, we can use the following depoissonization Lemma.
\begin{lemma}
When $n\to \infty$:
$$
p_n(\bz_0,\bz_1,t)\le p(n,\bz_0,\bz_1,t)(1+o(1)).
$$
\label{Lem:depoisson}
\end{lemma}
\begin{proof}
See Appendix~\ref{ap:depoi}.
\end{proof}

\subsubsection{Space-time Asymptotic Analysis}
\label{speedkernel}
We now evaluate the asymptotic behavior of the journey
density~$p_\nu(\bz_0,\bz_1,t)$. With a slight change of notation, we have substituted the node density instead of the number of nodes in the network, since in fact we are interested in the limit where $n$ tends to infinity, while $\nu=\frac{n}{\A}$ remains constant.
From Lemma~\ref{Lem:depoisson}, we see that we can equivalently evaluate the asymptotic behavior of the Poisson generating function coefficient $p(\nu,\bz_0,\bz_1,t)$ (where the number of nodes $n$ tends to infinity).

As we can observe by substituting the expressions of Lemma~\ref{Lem:laplace} in Corollary~3 (again with $\nu=\frac{n}{\A}$,
the asymptotic coefficient (corresponding to the journey density when $n\to\infty$) of the Poisson generating function
$\TW(\nu,\sigma)$, with $\sigma=(\zeta,\theta)$ a space-time vector, has a denominator $K(|\zeta|,\theta)$, such that (with $\rho=|\zeta|$):
$$K(\rho,\theta)=(1-\frac{n}{\A}\pi\frac{2}{\rho}I_1(\rho))\left(
\sqrt{(\tau+\theta)^2-\rho^2 v^2}-\tau\right)-\frac{n}{\A}4\pi v I_0(\rho).$$
The key of the analysis is the set $\CK$ of pairs
$(\rho,\theta)$ such that $K(\rho,\theta)=0$, called the {\it Kernel}. In fact any element of the Kernel (\ie a singularity of the Laplace transform) can be used to obtain an asymptotic estimate of the journey probability density.
We denote $(\rho_0,\theta_0)$ the element of the Kernel
that attains the minimum value $\frac{\theta}{\rho}$.
Notice that $(\rho_0,\theta_0)$ is a function of $\nu=\frac{n}{L^2}$.
We prove the following lemma.

\begin{lemma}
Let $\nu$ be fixed and $\theta_1>\theta_0$. There exists an $A_1$ such that,
when $|\bz|=|\bz_1-\bz_0|$ and $t$ both tend to infinity:
$$
p(\nu,\bz_0,\bz_1,t)\le A_1\exp(-\rho_0|\bz|+\theta_1 t)~.
$$
\label{Lem:saddle}
\end{lemma}

\begin{proof}
See Appendix~\ref{ap:asymptotic}.
\end{proof}

\subsubsection{Information Propagation Speed}

Let $\bz_0$, and $\bz_1$ be fixed.
Let $q_\nu(\bz_0,\bz_1,t)$ be the probability that there exists a journey that
arrives at distance less than 1 to a destination node at~$\bz_1$ before time~$t$.
\begin{lemma}
We have the upper bound:
$$
q_\nu(\bz_0,\bz_1,t)\le\int_{|\bz_1-\bz'|<1}p(\nu,\bz',\bz_1,t)d\bz'~.
$$\label{Lem:q}
\end{lemma}

\begin{proof}
By the definition of $q_\nu(\bz_0,\bz_1,t)$.
\end{proof}

Therefore, from Lemmas~\ref{Lem:depoisson},~\ref{Lem:saddle} and~\ref{Lem:q}, when $L\to\infty$ with $\bz_0$, $\bz_1$ fixed and
$\bz=\bz_1-\bz_0$, we
have the estimate $q_\nu(\bz_0,\bz_1,t)=O(\exp(\theta_1 t-|\bz|\rho_0))$ for
all $\theta_1>\theta_0$.

Clearly, $q_\nu(\bz_0,\bz_1,t)$ vanishes
very quickly when $t$ is smaller than the value such that
$-\rho_0|\bz|+\theta_0 t=0$, {\it i.e.}, when
$\frac{\theta_0}{\rho_0}=\frac{|\bz|}{t}$. This ratio
gives the upper bound for the propagation speed. In other words, point
$(\rho_0,\theta_0)$  achieves the lowest ratio
$\frac{\theta}{\rho}$ in the kernel set $\CK$. By expressing the kernel set $\CK$ using the function $K(\rho,\theta)$ from the previous section,
we obtain
Theorem~1.

\paragraph*{Remark}
We note that this result corresponds to the situation where all nodes speeds are of modulus $v$ (as assumed in Lemma~\ref{Lem:laplace}). Even if the speeds follow a different distribution, our analysis still applies, with the only change occurring in the Laplace transform of the motion vectors (but then the final form of Theorem~1 would be different). However, for an upper bound on the propagation speed, it suffices to
consider~$v$ as the maximum node speed.

To formally complete the proof, we need to address two remaining details: the contribution of the mirror images of the nodes (\ie to account for the nodes bouncing on the borders) and the destination's motion.
We note here that all node mirror images induce a contribution factor of order
$\exp(\theta_1 t-|\bz|\rho_0-x)$, where $x$ is the distance of the node from the border of the square network domain (see Appendix~\ref{ap:mirror}); for almost all nodes, $x$ is of the order of $L$, \ie the edge length of the square, which tends to infinity; therefore the contribution of the mirror images is negligible, since it decays exponentially in $\exp(-L)$.
For the destination's motion, it suffices to multiply the journey Laplace transform with the Laplace transform of the destination node excursion from its original position, to compute an upper bound on the propagation speed.
Similarly, the destination's motion also induces a negligible factor (see Appendix~\ref{ap:destination}).

\section{Sparse Two-Dimensional Networks}\label{sect:sparse}
\subsection{The Random Walk Model}
\paragraph*{Corollary~1}
When nodes move at speed $v>0$ in a random walk model
(with node direction change rate $\tau >0$), and when the square length
$L\to\infty$ but such that the node density $\frac{n}{\A}\to 0$, the propagation speed
upper bound is asymptotically equivalent to
$O(\sqrt{\frac{n v}{\A\tau}} v)$.

\begin{proof}
Let $(\rho,\theta)$ be an element of the set $\CK$. We have
$\theta=\sqrt{(\tau+\nu H(\rho))^2+\rho^2 v^2}-\tau$, with
$H(\rho)=\frac{4 \pi v I_0(\rho)}{1-\frac{n}{\A}\pi\frac{2}{\rho} I_1(\rho)}$. We have:
$H(\rho)=\frac{4 \pi v}{1-\pi\nu}+O(\rho^2),$
where $\nu=\frac{n}{\A}$. Therefore,
$$\theta=\sqrt{\tau^2+\rho^2 v^2}-\tau+\frac{\tau}{\sqrt{\tau^2+\rho^2 v^2}}H(\rho)\nu +O(\nu^2).$$
We obtain the ratio:
$$
\begin{array}{rcl}
\frac{\theta}{\rho}&=&\frac{H(\rho)\nu}{\rho}\frac{\tau}{\tau^2+\rho^2 v^2}+\frac{\sqrt{\tau^2+\rho^2 v^2}-\tau}{\rho}+O(\frac{\nu^2}{\rho})\\
&=&\frac{H(0)\nu}{\rho}+\frac{\rho v^2}{2\tau}+O(\frac{\nu^2}{\rho}+\nu\rho^2).
\end{array}
$$
Quantity $\frac{H(0)\nu}{\rho}+\frac{\rho v^2}{2\tau}$ is minimized with value
$v\sqrt{\frac{2\nu H(0)}{\tau}}$ attained at $\rho=\frac{\sqrt{2\nu\tau H(0)}}{v}$. Therefore
$\frac{\theta}{\rho}$ is minimized at value $v\sqrt{\frac{2\nu H(0)}{\tau}}+O(\nu^{3/2})$.
\end{proof}

\subsection{The Billiard Random Way-point Limit}
\label{Sect:RWP}

The billiard limit is equivalent to setting $\tau=0$.

\paragraph*{Corollary 2}
When nodes move at speed $v>0$ with $\tau=0$, and when
$L\to\infty$ but with node density $\nu=\frac{n}{\A}\to 0$,
the propagation speed upper bound is
$(1+ O(\nu^2))v$.

\begin{proof}
Now, the kernel set $\CK$ consists of the points $(\rho,\theta(\rho))$ where
$\theta(\rho)=v\sqrt{\rho^2+H_1(\rho)^2}$ with
$H_1(\rho)= \frac{\frac{n}{\A}4 \pi I_0(\rho)}{1-\frac{n}{\A}\pi\frac{2}{\rho} I_1(\rho)}$.
In this case, the
upper bound speed is proportional to $v$ with a factor of
proportionality equal to
$\sqrt{1+\left( \frac{H_1(\rho_0)}{\rho_0} \right)^2}$
where $\rho_0$
minimizes $\frac{H_1(\rho)}{\rho}$.
Since $\frac{H_1(\rho_0)}{\rho_0}=\frac{n}{\A}4 \pi \min\{\frac{I_0(\rho)}{\rho}\}+O(\nu^2)$, we get the estimate $(1+O(\nu^2))v$, proving Corollary~2.
\end{proof}

These corollaries are useful in order to see more intuitively the behavior of the upper bound of Theorem~1, when the node density is small, and consequently to better understand the fundamental performance limits of DTNs. Indeed, the case of sparse networks deserves special attention because of the potential applications and the necessity to use a delay tolerant architecture.
For instance, in the random walk model, it is important to notice that the information propagation speed diminishes with the square
root of the node density $\nu$. Furthermore, it is inversely proportional to the square root of the change of direction rate of the nodes (changing direction more frequently implies a smaller information propagation speed). In fact, the term in the square root in Corollary~1 is proportional to
the expected number of neighbors that a node meets during a random step.
Conversely, in the random way-point model, we notice that, surprisingly, the information propagation speed does not tend to~0 with the node density. In this case, the upper bound corresponds to the actual maximum speed of the mobile nodes (for instance, halving the node speed implies halving the information propagation speed).

\section{Multi-Dimensional Networks}
\label{Sect:multi}

In this section, we generalize our bounds on the information propagation speed when the network
map is in a space of dimension $D$, from $D=1$ to $D=3$.
This generalizes the case $D=2$ treated throughout the previous sections.

The network and mobility model is an extension of the unit disk model described in Section~\ref{Sect:model}. Again, we consider a network of $n$ nodes in a map of size
$\A=L^D$,
and we analyze the case where both $n,\A\to \infty$, such that the node density $\nu=\frac{n}{\A}$ tends to a (small) constant. Two nodes at distance smaller than $1$
can exchange information.
Initially, the nodes are distributed uniformly at random.
Every node follows an i.i.d. random trajectory, reflected on the
borders of the network domain like billiard balls.
The nodes change direction at Poisson rate~$\tau$ and keep a
uniform speed between direction changes, while the motion direction angles are
isotropic.

The journey decomposition as well as the asymptotic analysis in dimension $D=2$ can be directly generalized to other dimensions. We note here that the proofs of Lemmas~\ref{lem:card},~\ref{lem:fastest},~\ref{lem:lapl},~\ref{lm-poi},~\ref{lem:positive},~\ref{Lem:depoisson}, and~\ref{Lem:q} and Corollary~\ref{cor:laplace} hold independently of the network dimension.
Therefore, to analyze the propagation speed upper bounds in dimension $D$, we need to adapt
the journey Laplace transform expressions (Lemma~\ref{Lem:laplace}); we generalize the Laplace transforms in the following lemma.

\begin{lemma}
For $Y_D(\rho,\theta)$, $\Xi_D(\rho)$ and
$\Psi_D(\rho)$ defined (depending on $D$) in
Table~\ref{table:dim}, we have:
\begin{itemize}
\item $E(e^{-\zeta \cdot \bu})=\nu \Xi_D(\rho)$ when
$\bu$ is unitary and uniform on dimension $D$, with density $\nu$;
\item $E(e^{-\zeta \cdot \bu})=\nu \Psi_D(\rho)$ when $\bu$ is uniform
in the unit line, disk, ball (in dimensions $D=1,~2,~3$ respectively);
\item When all speeds are of modulus equal to $v$, we have
$E(e^{-\sigma \cdot M(\bv,\bw)})=\frac{1}{\frac{1}{Y_D(\rho,\theta)}-\tau}$.
\end{itemize}

\begin{table}[h]
\begin{center}
\begin{tabular}{|c|c|c|c|}
\hline
$D$&$Y_D(\rho,\theta)$&$\Xi_D(\rho)$&$\Psi_D(\rho)$\\
\hline
1 &$\frac{\tau+\theta}{(\tau+\theta)^2-\rho^2v^2}$&$2 \cosh(\rho)$&$2\frac{\sinh(\rho)}{\rho}$\\
\hline
2 &$\frac{1}{\sqrt{(\tau+\theta)^2-\rho^2v^2}}$&$2 \pi I_0(\rho)$&$\frac{2\pi}{\rho}I_1(\rho)$\\
\hline
3 &$ \frac{1}{2\rho v}\log\left(\frac{\tau+\theta+\rho v}{\tau+\theta-\rho v}\right)$&
$4 \pi \frac{\sinh(\rho)}{\rho}$&$\frac{4 \pi}{\rho^3}(\rho\cosh(\rho)-\sinh(\rho))$\\
\hline
\end{tabular}
\vskip 0.2cm
\caption{Definition of $Y_D(\rho,\theta)$, $\Xi_D(\rho)$ and $\Psi_D(\rho)$ (depending on $D$).}
\label{table:dim}
\end{center}
\end{table}

\label{Lem:laplaceD}
\end{lemma}
\begin{proof}
Equivalently to the proof of Lemma~\ref{Lem:laplace}, with $D-$dimensional integration.
\end{proof}

Moreover, we remark that the final result of the asymptotic analysis in Lemma~\ref{Lem:saddle} still holds in the case of networks in domains of dimension from $D=1$ to $D=3$. To adapt the proof, it suffices to substitute the respective Laplace transform expressions from Lemma~\ref{Lem:laplaceD} (see asymptotic analysis in the appendix in Section~\ref{ap:asymptotic}) and to compute the inverse Laplace transform in space dimension $D$ instead of dimension 2.
We can thus prove the following more general theorem.

\begin{theorem}
  In a network of $n$ nodes in a space of dimension $D$ and size $\A=L^D$, where $n\to\infty$ and $\A\to\infty$ such that the node density $\nu=\frac{n}{\A}$ remains constant, an upper bound of the information propagation speed is the smallest ratio $\min_{\rho,\theta>0}\left\{\frac{\theta}{\rho}\right\}$, with:
$$
\frac{1}{Y_D(\rho,\theta)} - \tau - \frac{2 v \nu \Xi_D(\rho)}{1-\nu \Psi_D(\rho)}=0,
$$
where $v$ is the maximum node speed, $\tau$ the node direction change rate, and the values of $Y_D(\rho,\theta)$, $\Xi_D(\rho)$ and
$\Psi_D(\rho)$ are defined (depending on $D$) in
Table~\ref{table:dim}.
\label{Theo:dim}
\end{theorem}

\paragraph*{Remark} From the definition of $\Psi_D()$, the previous expression has meaning
when $\nu<\frac{1}{V_D}$, where $V_D$ is the ``volume'' of transmission radius $1$: $V_D=2$ in $1-$D,
$V_D=\pi$ in $2-$D, and $V_D=\frac{4 \pi}{3}$ in $3-$D.
Above this density threshold,
the upper bound for the information propagation speed is infinite.
Such a behavior is expected in dimensions 2 and 3, since it is known that there exists a critical density above which the network graph percolates, \ie there exists an infinite connected component.
However, a tighter analysis in dimension 1 would yield a propagation speed increasing exponentially with the node density, in accordance with the size of the largest connected component.

\begin{proof}
Initially, we consider a fixed destination; however, we note that the discussion of the moving destination in the appendix (Section~\ref{ap:destination}) is valid in other dimensions too, therefore the propagation speed upper bound remains unchanged if the destination moves as the other nodes.

Using the new Laplace transforms of Lemma~\ref{Lem:laplaceD},
the asymptotic coefficient of the Poisson generating function
$\TW(\nu,\sigma)$ (defined in Corollary~\ref{cor:laplace}), with $\sigma=(\zeta,\theta)$ a space-time vector, has a denominator $K_D(|\zeta|,\theta)$, such that (with $\rho=|\zeta|$):
$$K_D(\rho,\theta)=\frac{1}{Y_D(\rho,\theta)} - \tau - \frac{2 v \nu \Xi_D(\rho)}{1-\nu \Psi_D(\rho)}.$$

The set $\CK$ of pairs
$(\rho,\theta)$ such that $K_D(\rho,\theta)=0$ corresponds to the new Kernel set.
Therefore, from the new Kernel expression and Lemmas~\ref{Lem:saddle} and~\ref{Lem:q}, we obtain the expression for the information propagation speed upper bound: the smallest ratio $\frac{\theta}{\rho}$, with $K_D(\rho,\theta)=0$.

Again, to complete the proof, we must account for the fact that the nodes bounce on the network domain borders, \ie to add the contributions of the node mirror images as discussed in Section~\ref{Sect:model}, in an infinite domain of dimension $D$.
According to the analysis of the two-dimensional case (see Appendix~\ref{ap:mirror}), the contribution of the mirror images is negligible (in dimension 1 it suffices to consider only the closest mirror image, while in dimension 3, we must consider the 4 closest images).
\end{proof}

\section{Slowness of Information Propagation Plots}
\label{sect:slowness}

\begin{figure*}[t1]
\begin{center}
\begin{minipage}[b]{0.49\linewidth}
\centering
{\includegraphics[width =5.6cm]{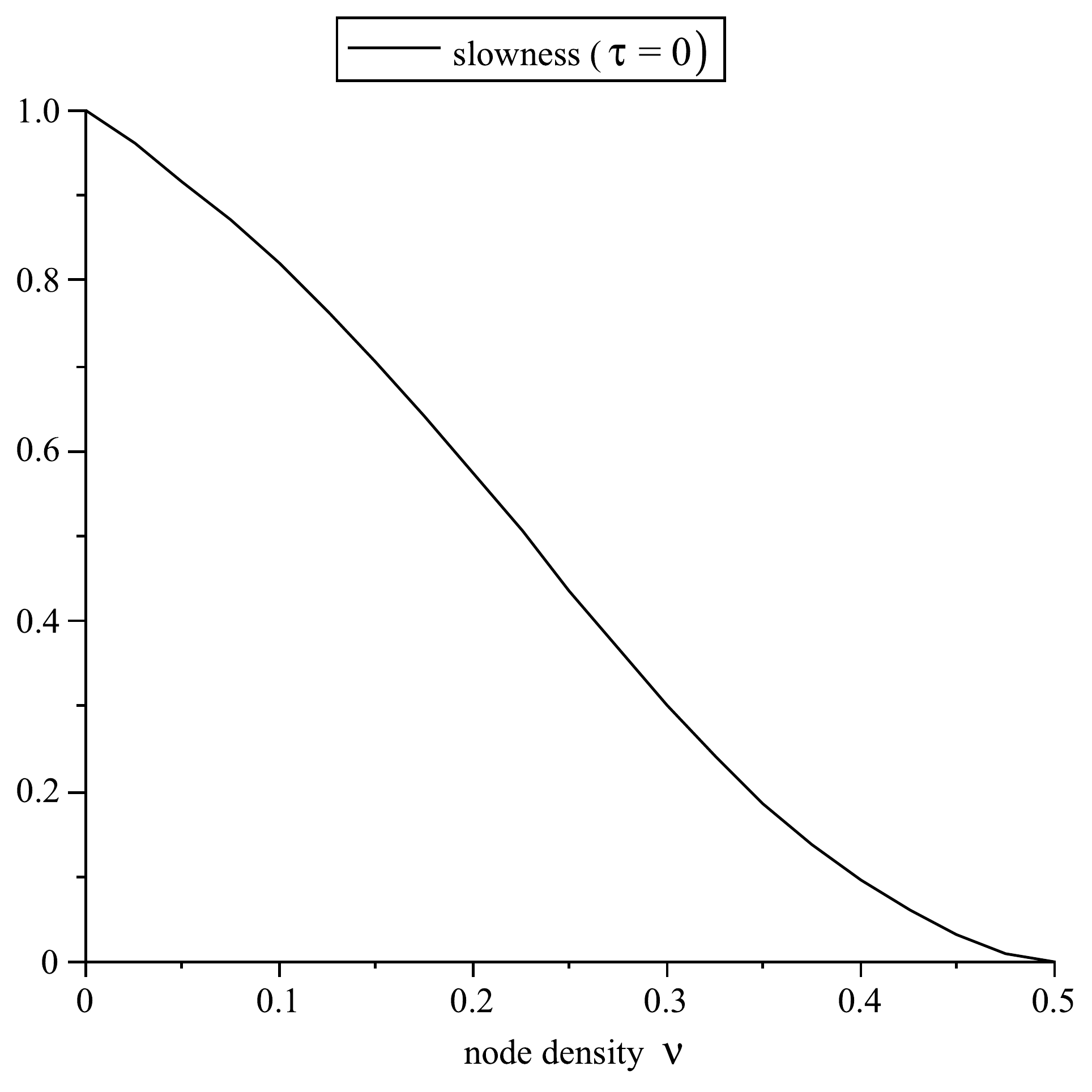}}
\end{minipage}
\begin{minipage}[b]{0.49\linewidth}
\centering
{\includegraphics[width =5.6cm]{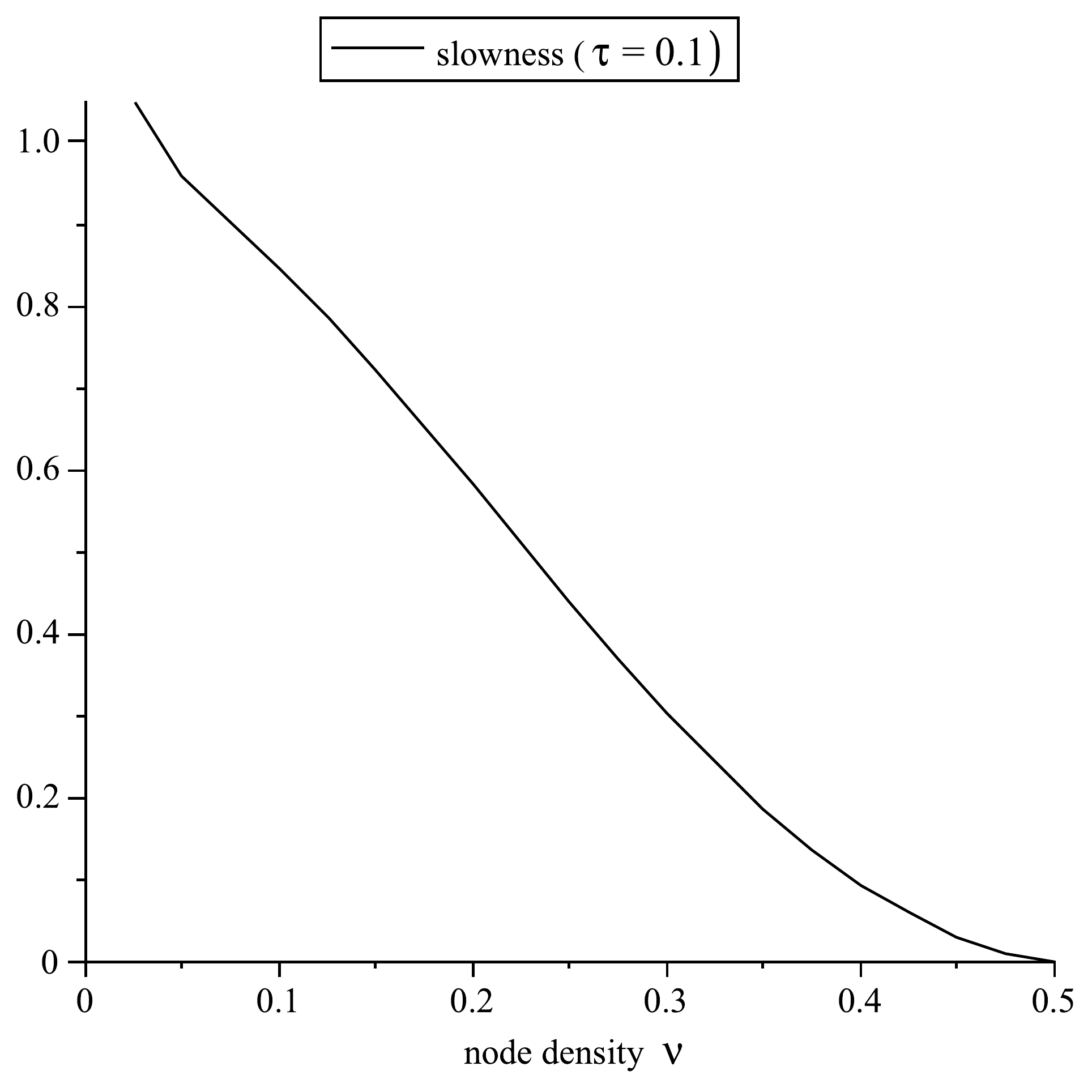}}
\end{minipage}
\caption{Theoretical lower bound of slowness versus mobile node density $\nu$ when $\tau=0$ (left) and $\tau=0.1$ (right), in $1-$D networks.}
\label{slownessD1}
\end{center}
\end{figure*}

\begin{figure*}[t!]
\begin{center}
\begin{minipage}[b]{0.49\linewidth}
\centering
{\includegraphics[width =5.6cm]{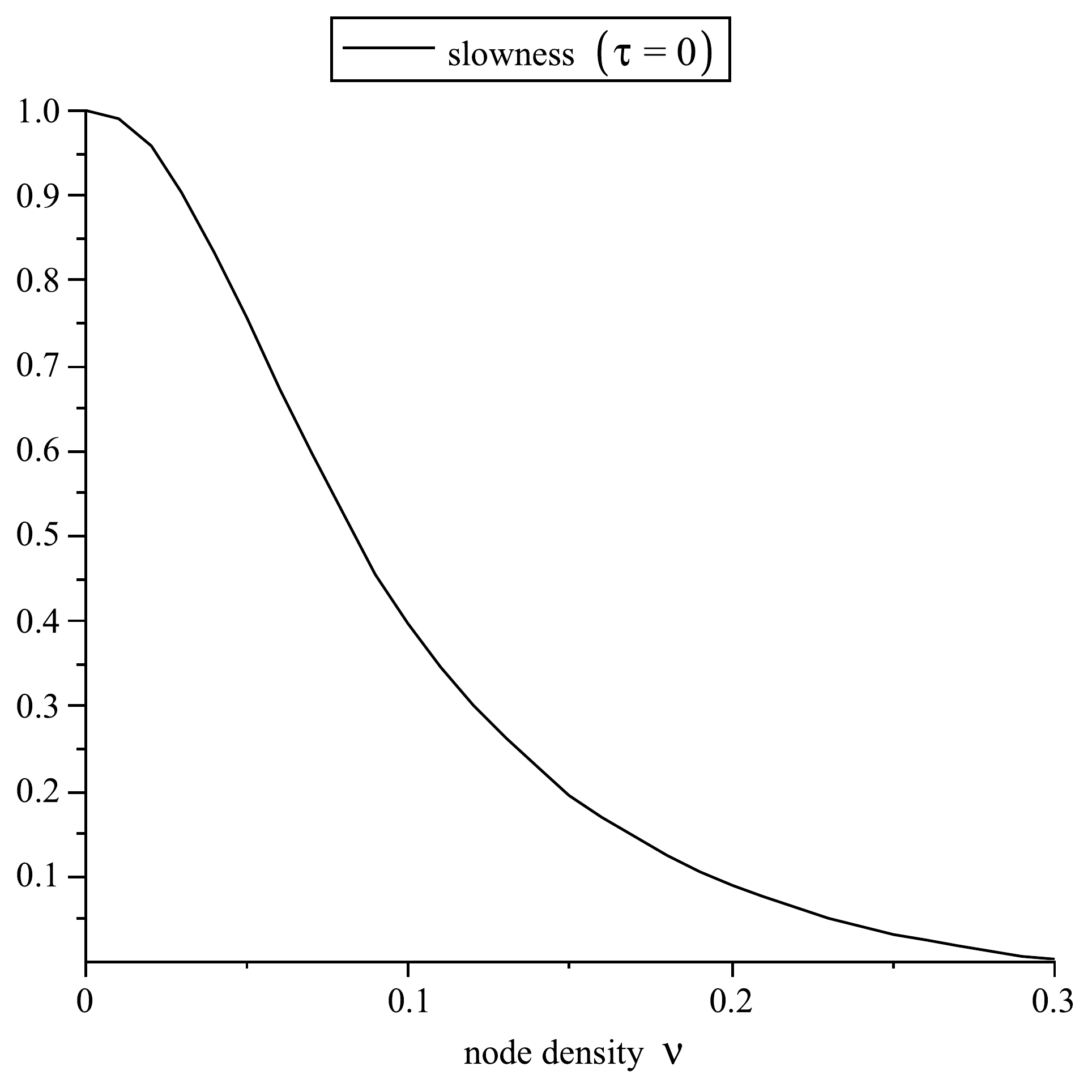}}
\end{minipage}
\begin{minipage}[b]{0.49\linewidth}
\centering
{\includegraphics[width =5.6cm]{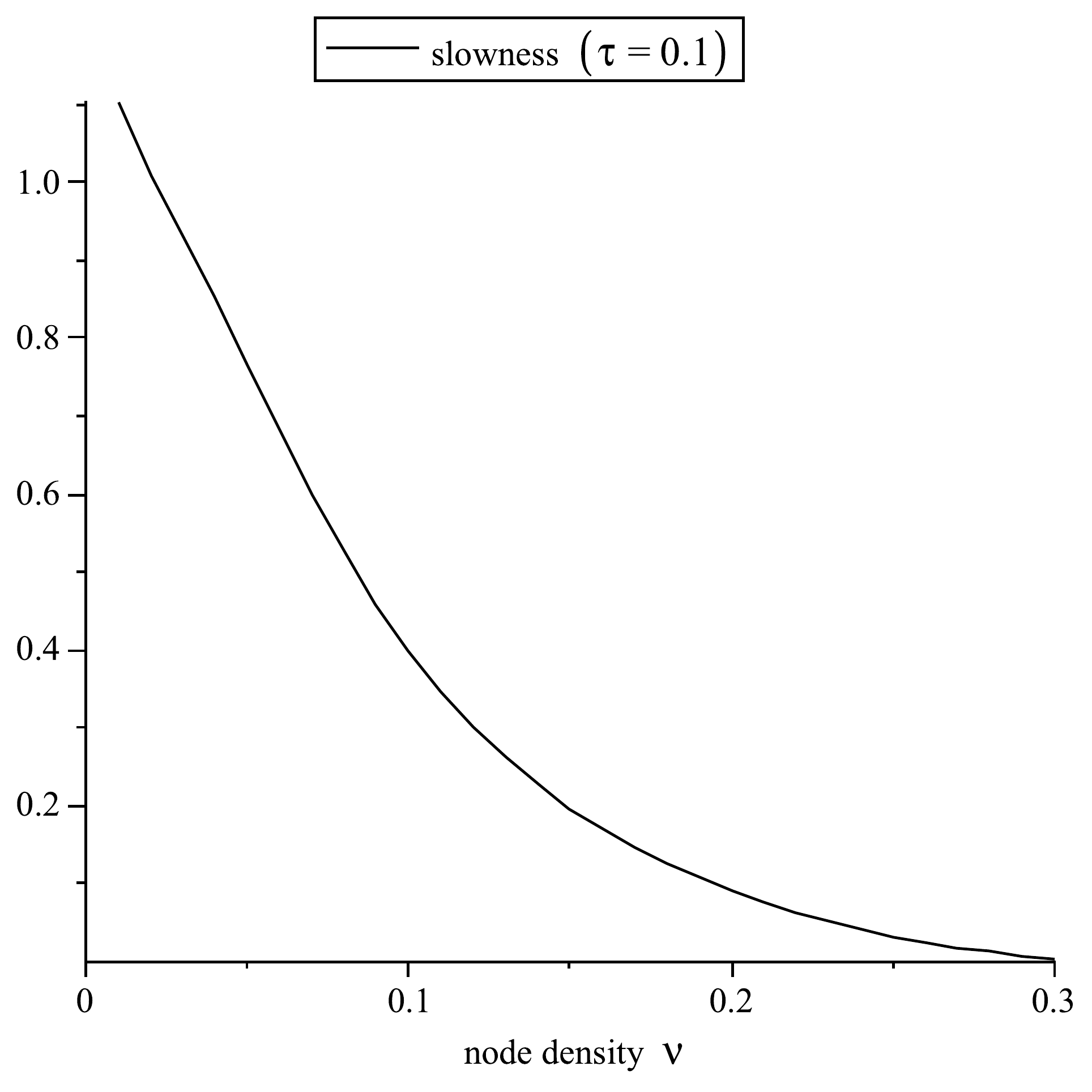}}
\end{minipage}
\caption{Theoretical lower bound of slowness versus mobile node density $\nu$ when $\tau=0$ (left) and $\tau=0.1$ (right), in $2-$D networks.}
\label{slowness00}
\end{center}
\end{figure*}

\begin{figure*}[t!]
\begin{center}
\begin{minipage}[b]{0.49\linewidth}
\centering
{\includegraphics[width =5.6cm]{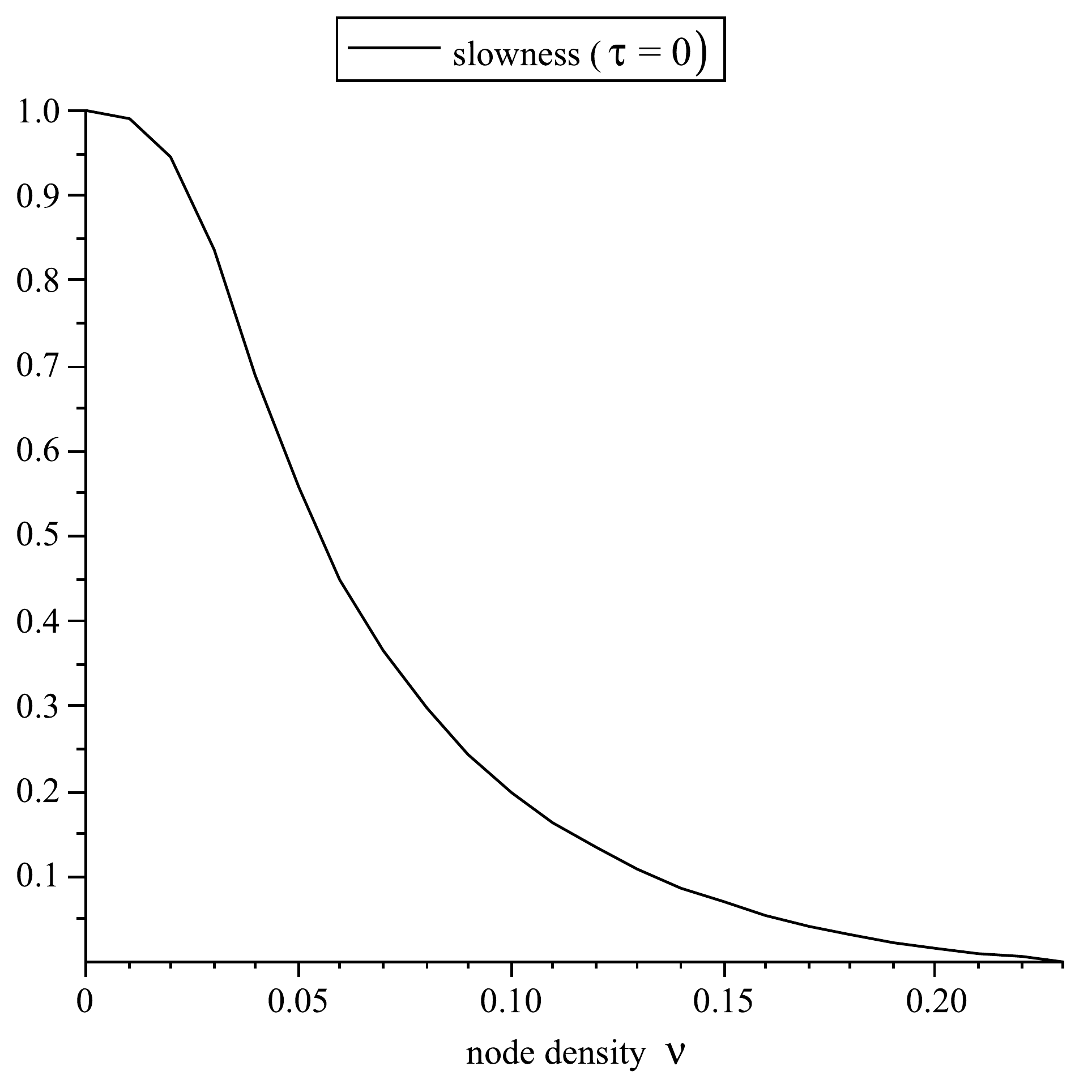}}
\end{minipage}
\begin{minipage}[b]{0.49\linewidth}
\centering
 {\includegraphics[width =5.6cm]{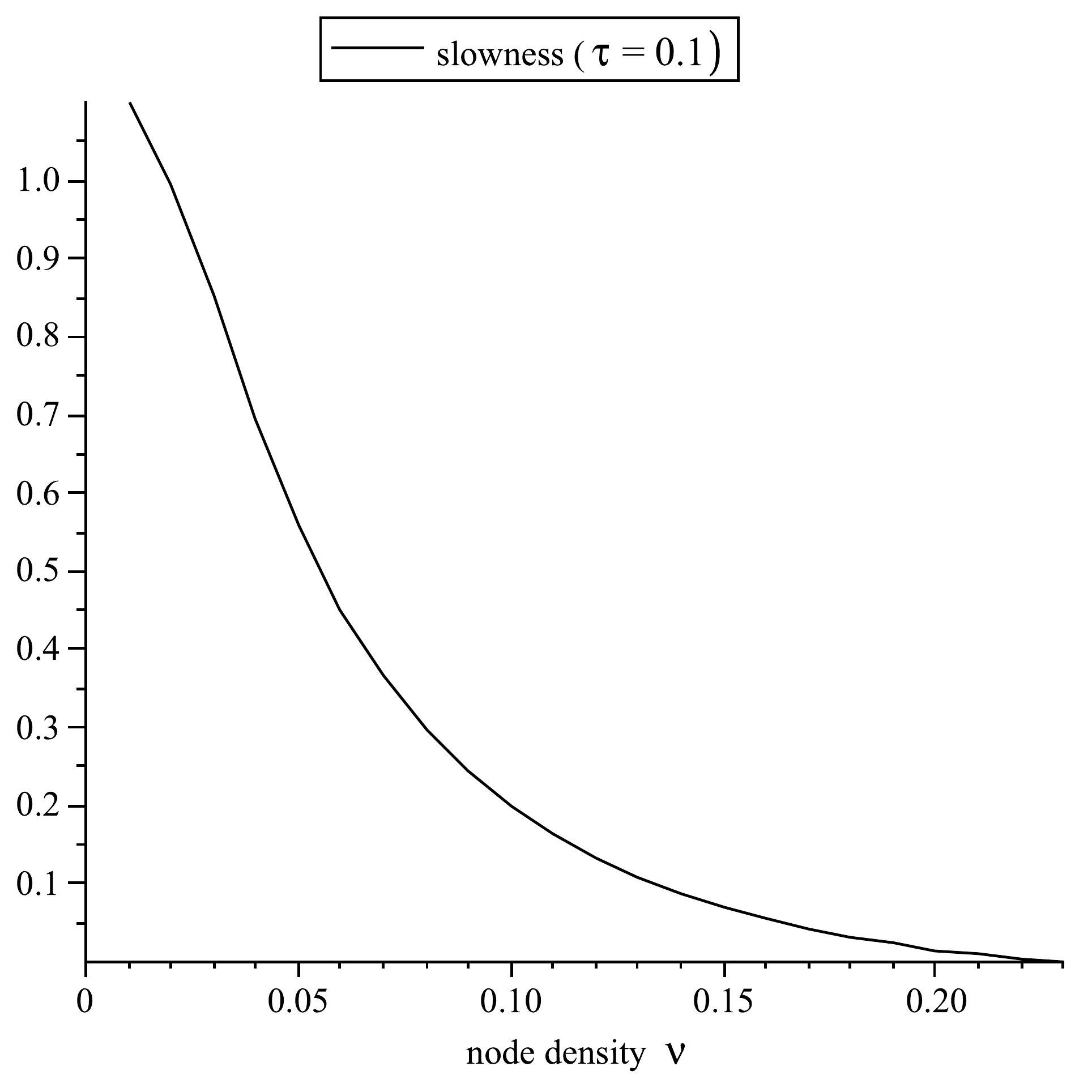}}
\end{minipage}
\caption{Theoretical lower bound of slowness versus mobile node density $\nu$ when $\tau=0$ (left) and $\tau=0.1$ (right), in $3-$D networks.}
\label{slownessD3}
\end{center}
\end{figure*}

To illustrate the behavior of the upper bound for the information
propagation speed when the mobile density $\nu$ varies, we define the
{\em slowness}, \ie the inverse of the information propagation speed,
for which our theoretical study now provides lower bounds.
Plotting results (obtained by
numerical resolution of Theorem~2, or, equivalently, of Theorem~1 in the two-dimensional case) of our 
lower bounds are presented in
Figures~\ref{slownessD1},~\ref{slowness00} and~\ref{slownessD3}, in networks of 1,~2 and~3 dimensions respectively,
where we consider a unit maximum node speed: $v=1 ms^{-1}$.

Interestingly, in all dimensions, the limit of the information propagation speed when the node density tends to zero corresponds to the maximum node speed in the billiard mobility model ($\tau=0$), while the propagation slowness is unbounded for small node densities in the random walk model (\ie when $\tau>0$, the information propagation speed diminishes with the node density).

We remark that the slowness drops to 0 at $\nu=1/V_D$, with $V_D=2$ in $1-$D,
$V_D=\pi$ in $2-$D, and $V_D=\frac{4 \pi}{3}$ in $3-$D: this
corresponds to the limit of our model. Recall, that this is a lower
bound of the slowness (equivalent to the upper bound for the propagation
speed). The actual slowness should continue to be non-zero beyond
$\nu=1/V_D$.

Furthermore, in the two-dimensional case (Figure~\ref{slowness00}), we notice that the slowness
is in $1-O(\nu^2)$ for the billiard - random way point limit (\ie $\tau=0$), confirming Corollary~2; for the random walk,
we notice that the slowness is unbounded when $\nu \to 0$, confirming
the $O(\frac{1}{\sqrt{\nu}})$ theoretical behavior proved in
Corollary~1.

\section{Simulations}
\label{Sect:sim}

In this section,
we evaluate the accuracy of our theoretical upper bound in
different scenarios by comparing it to the average information
propagation time obtained by simulating a full epidemic broadcast in a two-dimensional network (as described
in Section~\ref{Sect:intro}).
For all the simulations, we use a unit-disk graph model (\ie a radio range of $1m$), and the mobile node
speed is $1 ms^{-1}$.
Two commonly used mobility models are simulated:
the random way-point model (which corresponds to our setting $\tau=0$ as
described in Subsection~\ref{Sect:RWP}) and the random walk model (which
corresponds to our setting $\tau=0.1$).
We study the two mobility models (Figure~\ref{rdwp} for billiard random way-point mobility,
and Figure~\ref{rdwk} for random walk mobility)
for different node density~$\nu$
and area values ($\nu=0.025$ on a $80\times 80$ square, $\nu=0.05$ on
a $60\times 60$ square, $\nu=0.1$ on a $40\times 40$ square,
respectively).

In Figures~\ref{rdwp} and~\ref{rdwk},
we depict the simulated propagation time versus the distance (plots), and we compare it to the
theoretical bound, \ie a line of fixed slope (in green - bottom). The slope is obtained from the analysis
in Section~\ref{Sect:analysis}; it represents the
slowness illustrated for corresponding density values in Figure~\ref{slowness00}.
In the figures, time is measured in seconds, and distance in meters, therefore, the inverse slope of the plots provides us with the information propagation speed in~$ms^{-1}$.
What is important is the comparison of the slopes at infinity. We notice that the measurements very quickly converge to a straight line of fixed slope, which implies a fixed information propagation speed.

Simulations show that the theoretical slope is clearly a lower
bound on the slowness, as proved in Theorem~1.
We also compare the simulation measurements with a second line of fixed slope (red - top). This line is provided only for comparison and corresponds to the heuristic situation where we assume that node movements and emissions are completely independent (according to the framework of~\cite{JM07} in an infinite network). Interestingly enough, the simulations show that the heuristic
bound provides an accurate slope (the theoretical slope we provide is smaller, since in order to prove a rigorous bound on the information propagation speed, we work with an upper bound on the journey probability density for the journey decomposition in Section~\ref{sect:dependent}).

\section{Conclusion}
\label{Sect:conclusion}

In this paper,
we have initiated a characterization of the
information propagation speed of Delay Tolerant mobile
Networks (DTNs)
by providing a theoretical upper bound for large scale but finite two-dimensional networks (Theorem~\ref{Theo:upper} and Corollaries~1 and~2) and multi-dimensional networks (Theorem~\ref{Theo:dim}).
Such theoretical bounds are useful in order to increase our understanding
of the fundamental properties and performance limits of DTNs, as well as to evaluate and/or optimize the performance of specific routing algorithms.
The model used in our analytical
study is sufficiently general to encapsulate many popular mobility
models (random way-point, random walk, Brownian motion).
We also performed simulations
for several scenarios to show the validity of our
bounds.

Our methodology and space-time journey analysis provide a general framework for the derivation of analytical bounds on the information propagation speed in DTNs. Therefore, future investigations should consider extending the
analysis to other neighboring models different from
unit disk graphs (\eg quasi-disk graphs, probabilistic models), proving tighter bounds (\eg similar to the heuristic bound we discussed in the simulations section), and generalizing to other mobility models or comparing the results with real traces. Another interesting direction for further research would be to compare the implications of our analysis on the delay of common routing schemes, such as epidemic routing, with the results presented in previous work on DTN modeling~\cite{GNK05,ZNKT07}, under the frequently used assumption that the inter-meeting time between pairs of nodes follows an exponential distribution.

\appendix

\subsection{Node Meeting Rate}\label{ap:rate}
We consider two nodes moving at speeds $\bw_1$ and $\bv_2$, respectively.
We compute
the average rate at which the second node
enters
the neighborhood range of the previous node on the radius
$\bu_1$ (the vector $\bu_1$ is centered on the position of the first node and is of modulus $|\bu_1|=1$).
The relative speed of the nodes is $\bw_1-\bv_2$.
The projection of the relative speed on the vector $\bu_1$ equals $\left(\bu_1\cdot (\bw_1-\bv_2)\right)\bu_1$.
If the dot product is positive, the rate at which the second node enters the neighborhood range of the first node at
$\bu_1$, equals $\bu_1\cdot (\bw_1-\bv_2)\frac{1}{\A}$, where quantity $\frac{1}{\A}$ is the density of presence of the second node.
On the other hand, if the dot product is negative, the nodes move in such directions that they cannot meet on the radius
$\bu_1$.

\subsection{Proof of Lemma~\ref{Lem:laplace} (Laplace Transform Expressions)}\label{ap:laplace}
\begin{itemize}
\item The expression $E(e^{-\zeta \cdot \bu})$, when $\bu$ is uniform on the unit circle, with density $\nu$,
is equal to $\nu \int_0^{2\pi}e^{|\zeta|\cos\phi}d\phi
=2 \pi \nu I_0(|\zeta|)$.
\item The expression $E(e^{-\zeta \cdot \bu})$ when $\bu$ is uniform on the unit disk, with
density $\nu$, is
equal to: $$\nu\int_0^{2\pi}d\phi\int_0^1e^{-r|\zeta|\cos\phi}2\pi rdr.$$
In
Taylor series, we have $\nu\pi\sum_k(\frac{|\zeta|}{2})^{2k}\frac{1}{(k+1)!k!}$,
which in turn is equal to $\nu\pi\frac{2}{|\zeta|}I_1(|\zeta|)$.
\item We define a carry segment $S_c(\bv)$ as the space-time vector corresponding to a node motion of constant speed $\bv$, until the node changes direction.
Since nodes change direction at Poisson rate $\tau$ and the speed modules is $v$, we have the expression (with $\sigma = (\zeta,\theta)$): $E(e^{-\sigma \cdot S_c(\bv)})=\tau Y(\zeta,\theta)$, where $Y(\zeta,\theta)=\int_0^\infty e^{|\zeta| v t-\theta t}e^{-\tau t}dt=\frac{1}{\sqrt{(\theta+\tau)^2-|\zeta|^2 v^2}}$ (see also~\cite{JM07}).
The node motion vector $M(\bv,\bw)$ corresponds to an arbitrary sequence of carry segments and a final segment,
which ends with the reception of the information packet by the final destination, instead of a change of direction.
Therefore, we have the following simple expression inspired from
combinatorial analysis:
$E(e^{-\sigma \cdot M(\bv,\bw)})=\frac{Y(\zeta,\theta)}{1-\tau Y(\zeta,\theta)}$.
This is the equivalent of the formal identity $\frac{1}{1-y}=1+y+y^2+y^3+...$, which represents the Laplace transform of an arbitrary sequence of random variables with Laplace transform $y$ (\emph{cf.}~\cite{FS08}), while the term in the numerator corresponds to the final motion vector before emitting to the destination (therefore, this term is not multiplied by the direction change rate~$\tau$).
\end{itemize}

\subsection{Proof of Lemma~\ref{Lem:depoisson} (Depoissonization)}\label{ap:depoi}
By Cauchy integration (\emph{cf.}~\cite{js98}):
$$
p_n(\bz_0,\bz_1,t)=\frac{n!}{2i\pi}\oint p(Z,\bz_0,\bz_1,t)\frac{e^Z}{Z^{n+1}}dZ,
$$
where the integration loop encircles the origin on the complex plane.
We take the circle of center 0 and radius $n$:
$$
p_n(\bz_0,\bz_1,t)=\frac{n!}{2\pi}\int_0^{2\pi} p(ne^{i\phi},\bz_0,\bz_1,t)
\frac{\exp(n e^{i\phi})}{e^{in\phi}}d\phi.
$$
Therefore,
$$
\begin{array}{rcl}
p_n(\bz_0,\bz_1,t)&\le&p(n,\bz_0,\bz_1,t)\frac{n!}{2\pi n^n}\int_0^{2\pi}
e^{n\cos\phi}d\phi\\
&=&p(n,\bz_0,\bz_1,t)\frac{n!}{ n^n}I_0(n).
\end{array}
$$
Using Stirling and Bessel asymptotics, $n!=\sqrt{2\pi n}n^ne^{-n}(1+o(1))$
and $I(n)=\frac{1}{\sqrt{2\pi n}}e^n(1+o(1))$, we complete the proof.

\subsection{Proof of Lemma~\ref{Lem:saddle} (Asymptotic Analysis)}\label{ap:asymptotic}
We prove this lemma for a random mobility model with speed of constant modulus, but generalization is straightforward.
In this case, with $\sigma=(\zeta,\theta)$ a space-time vector:
$
\TW(\nu,\sigma)=\frac{1}{K(|\zeta|,\theta)}
$,
with $K(|\zeta|,\theta)=(1-\nu\pi\frac{2}{|\zeta|} I_1(|\zeta|)) \left( \sqrt{(\theta+\tau)^2-|\zeta|^2v^2}-\tau -2\pi v\nu I_0(|\zeta|) \right)$.

Notice that $\frac{1}{K(|\zeta|,\theta)}$
is in fact an analytic function of $|\zeta|^2$ and
of $\theta$ with non negative coefficients.
One must be aware that the quantity $|\zeta|^2$ refers to the sum
of the square of the coefficients of $\zeta$ and not to the sum of the square of their
modulus, and therefore induces an analytical function.
Thus, the definition domain of
$\TW(\nu,\sigma)$ contains all tuples $\sigma=(\zeta,\theta)$ such that~$(|\zeta|,\theta)$ belongs to the definition domain of
$\frac{1}{K(|\zeta|,\theta)}$.

In particular,
if there exists a real tuple $(\rho_1,\theta_1)$ in the
Kernel~$\CK$, then the tuples $(\rho,\theta)$
such that $\Re(\theta)>\theta_1$
and $\Re(\rho)\le\rho_1$ belong to the definition domain of
$\frac{1}{K(\rho,\theta)}$.

Due to the asymptotic~\cite{AS65} on modified Bessel functions:
$I_\alpha(\rho)\sim\sqrt{\frac{2i}{\pi\rho}}
\cos\left(\frac{\rho}{i}-\frac{\alpha\pi}{2}-\frac{\pi}{4}\right)$, we
can enlarge the definition domain of $\frac{1}{K(\rho,\theta)}$ to
$Re(\rho)\le\rho_1+\frac{1}{2}\log|\Im(\rho)|$, when $\Im(\rho)$ is
large.

Since the Laplace transform of space-time density $p(\nu,\bz,t)$ is
$\frac{\TW(\nu,\sigma)}{\theta}$ (the denominator $\theta$ comes from the
fact that we cumulate all journeys arriving at $\bz$ within $t$),
$p(\nu,\bz,t)$ can be expressed via the inverse
Laplace transform:
$$
p(\nu,\bz,t)=
(\frac{1}{2i\pi})^3\int_{\Re(\theta)=\theta_1}\-\int_{\Re(\zeta)=\zeta_1}
\TW(\nu,\sigma)
e^{\zeta\cdot \bz+\theta t}
d\zeta\frac{ d\theta}{\theta},
$$
where $(\zeta_1,\theta_1)$ is any real element of the definition domain.

For our purpose, we will directly take
$\theta_1>\theta_0$ and $\zeta_1=-\frac{\rho_0}{|\bz|}\bz$.
Notice that the integration domain in $\zeta$ is an imaginary plane and in $\theta$
an imaginary axis (\ie integration is 3-dimensional); this explains the cubic
factor $\frac{1}{(2i\pi)^3}$.

Without loss of generality, we assume that $\bz$ is co-linear
with the first axis: {\it i.e.}, $\bz=(x,y)$ and $y=0$. We denote
$\zeta=(\zeta_x,\zeta_y)$.
The integral in $\theta$ absolutely converges because it is in
$\frac{1}{\theta}^2$,
but we cannot conclude the same for the absolute convergence of
$\TW(\zeta,\rho)$ which converges only in $\frac{1}{|\zeta|}$ when
$\Im(\zeta)$ tends to infinity.
To this end, we move the integration surface of $\zeta$ in a suitable
way that will lead to an absolute convergence. This move
is made possible by the multi-dimensional analytical nature of the
functions.

We first define a function $\rho(a)=\rho_0+ia$ for $|a|<B$ for some $B>0$ and
$\rho(a)=\rho_0+ia+\frac{1}{2}\log(|a|)$ for
$|a|>B$, such that $(\rho(a),\theta_1)$ always belong to the
definition domain of $\frac{1}{K(\rho,\theta)}$.
Second, we define the new surface of integration for $\zeta$ as the
the union of Minkowski hyperbolic sections defined by
$M(a)=\{\zeta,|\zeta|=\rho(a),\exists b>0:\Re(\zeta)=b\zeta_1\}$.
In other words $\Re(\zeta_y)=0$.
From the identity $\zeta_x^2=\rho(a)^2+(\Im(\zeta_y))^2$, we
get $|\Re(\zeta)|\ge\max\{\Re(\rho(a)),|\Im(\zeta)|\}$. Therefore,
$$
\begin{array}{rcl}
|\int_{M(a)}\exp(-\bz\cdot \zeta)d\zeta|&\le&\int_{M(a)}\exp(-|\bz|\Re(\zeta))|d\zeta|\\
&\le&\frac{\exp(-\Re(\rho(a))|\bz|)}{|\bz|}.
\end{array}
$$
Since $\exp(-\Re(\rho(a))|\bz|)=e^{-\rho_0|\bz|}(a^{-|\bz|})$, for $|a|>B$,
the integral converges absolutely in $a$ when we add the contribution
of all Minkowski sections, as soon as $|\bz|>2$.

\subsection{Contribution of Mirror Images}\label{ap:mirror}
Let us denote $p(\nu,\bz,t)$ as the density distribution of the journeys that connect
two points $\bz_0$ and $\bz_1$ in the original square which delimits the network domain, such that $\bz_1-\bz_0=\bz$,
with the journey starting at time $0$ and arriving at the destination before $t$. Notice that the
journey does not connect to the mirror images of $\bz_1$.
To account for the mirror images,
when $\bz_1=(x_1,y_1)$,
we need to add the three closest
images at: $\bz_1+2(L-x_1)e_x$,
$\bz_1+2(L-y_1)e_y$ and $\bz_1+2(L-x_1)e_x+2(L-y_1)e_y$
with $e_x=(1,0)$ and $e_y=(0,1)$. Adding all possible periodic mirror images,
we get the identity:
$$
\begin{array}{l}
p(\nu,\bz_0,\bz_1,t)=p(\nu,\bz,t)\\
~~~+p(\nu,\bz+2(L-x_1)e_x,t)+p(\nu,\bz+2(L-y_1)e_y,t)\\
~~~+p(\nu,\bz+2(L-x_1)e_x+2(L-y_1)e_y,t)\\
~~~+\sum_{(j,k) \neq(0,0)} p(\nu,\bz+2jLe_x+2kLe_y,t)\\
~~~~~+p(\nu,\bz+2(L-x_1)e_x+2jLe_x+2kLe_y,t)\\
~~~~~+p(\nu,\bz+2(L-y_1)e_y+2jLe_x+2kLe_y,t)\\
~~~~~+p(\nu,\bz+2(L-x_1)e_x+2(L-y_1)e_y+2jLe_x+2kLe_y,t).
\end{array}
$$

The dominant terms in the expression of $p(\nu,\bz_0,\bz_1,t)$ in addition to $p(\nu,\bz,t)$ correspond to the three closest images. Since we have shown in the previous section that $p(\nu,\bz,t)$ decreases exponentially with $\bz$, from the above identity, the additional factor induced by the mirror images of a given node is of order $\exp(\theta_1 t-|\bz|\rho_0-x)$, where $x$ is the distance of the node from the border of the square network domain.

\subsection{Moving Destination}\label{ap:destination}

We consider that
the destination can move as the other nodes, starting at position
$\bz$ at time $t=0$.
We show that
the asymptotic propagation speed upper bound does not change when
$(\bz,t)$ tend to infinity.

For this end, it suffices to multiply the journey
Laplace transform with the Laplace transform of the
node excursion from its original position.  The excursion Laplace
transform is obtained from the motion Laplace transform $E(e^{-\sigma \cdot M(\bv,\bw)})$, and it has
the expression
$\frac{1}{\frac{1}{Y(\zeta,\theta)}-\tau}$,
where $Y(\zeta,\theta)$ is defined as previously in the Laplace transform calculations (see the proof of Lemma~\ref{Lem:laplace} in the two-dimensional case, and the definition of $Y_D(\zeta,\theta)$ in Lemma~\ref{Lem:laplaceD} for the multidimensional case).
Therefore, the Poisson generating function $\TW^\ast (Z,\sigma)$ of the new Laplace
transform  equals $\TW(Z,\sigma)\frac{1}{\frac{1}{Y(\zeta,\theta)}-\tau}$ (with $\TW(Z,\sigma)$ the Poisson generating function corresponding to a fixed destination, defined in Corollary~\ref{cor:laplace}).
The function $\TW^\ast (Z,\sigma)$ has two sets of poles, the set $\CK$ (described in Section~\ref{speedkernel}) and the new set $\CK'$
corresponding to the set $\{(\rho,\theta):\theta=\rho v-\tau\}$, \ie the roots of the denominator of the excursion Laplace transform. The
last set is dominated on the right by $\CK$: for all
$(\rho,\theta')\in\CK'$, there is a $(\rho,\theta)$ in $\CK$ with
$\theta>\theta'+B$ and $B>0$.  Hence, the contributions from $\CK'$ will be
exponentially negligible (of order $\exp(-Bt)$) compared to the main
contribution from $\CK$, and the propagation speed upper-bound does not change from the value computed in Section~\ref{sect:speed}.

\clearpage
\begin{twocolumn}
\begin{figure}[t]
\begin{center}
\fbox{\includegraphics[width =7cm]{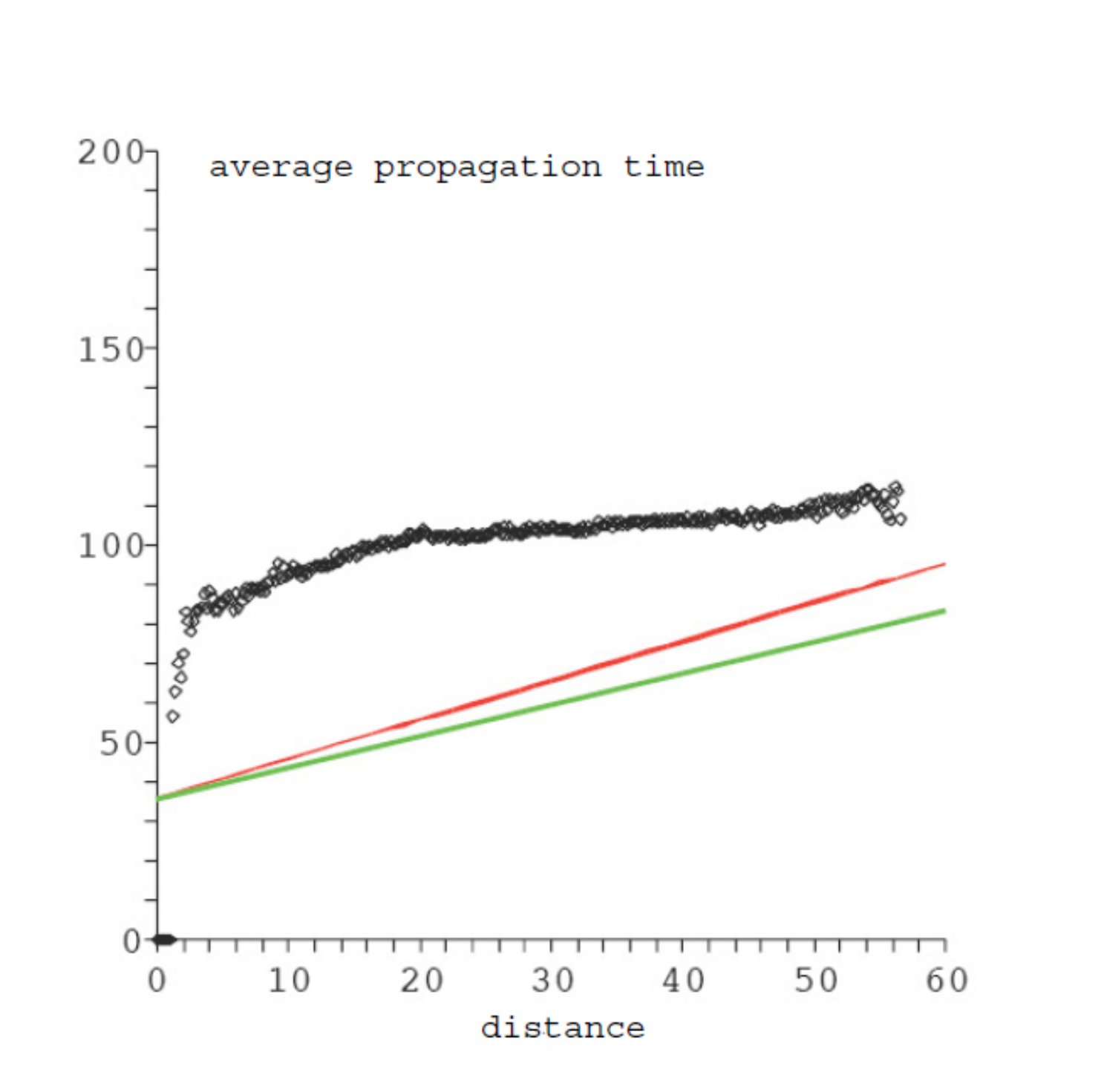}}
\vskip 0.6cm
\fbox{\includegraphics[width =7cm]{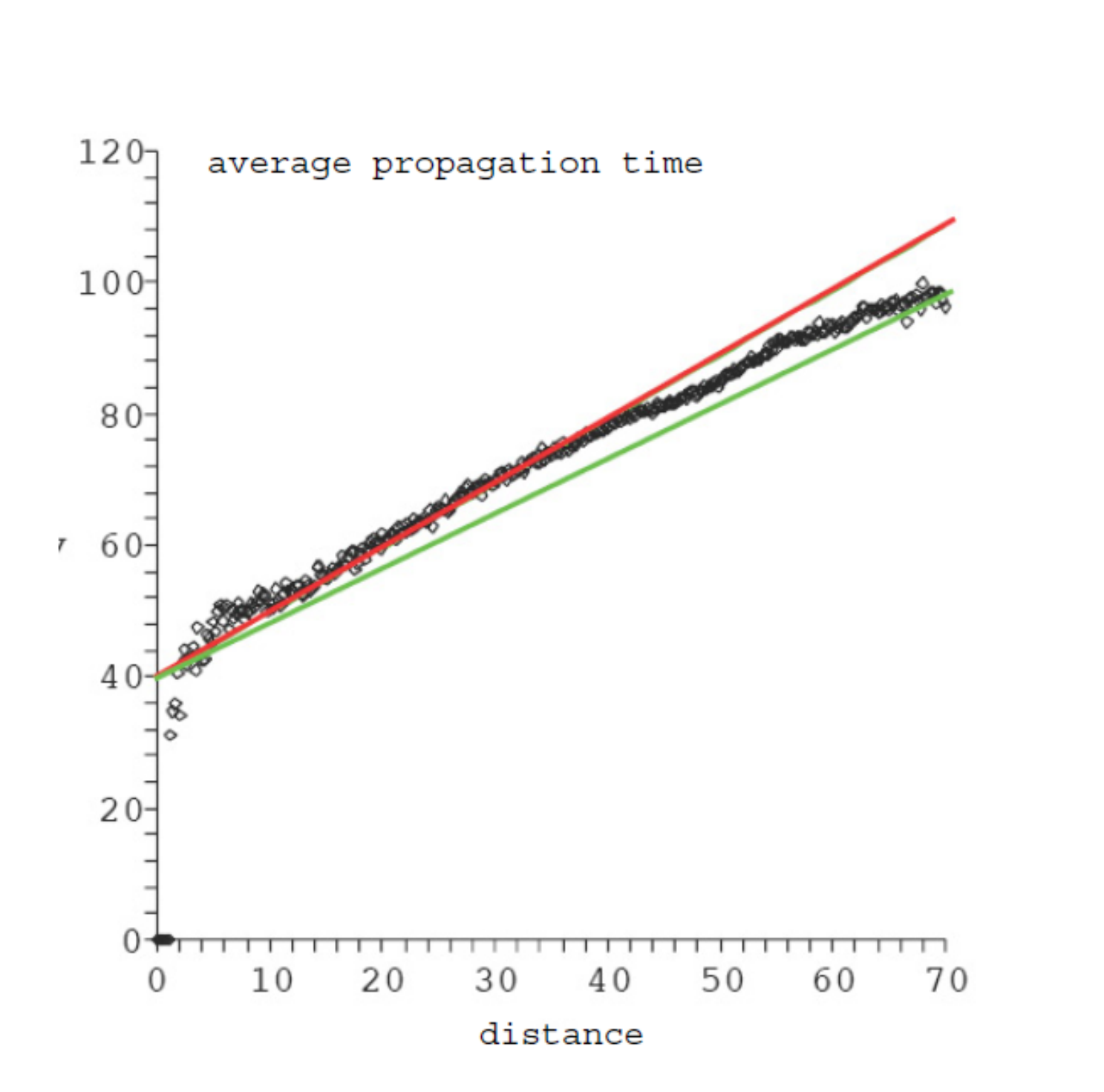}}
\vskip 0.6cm
\fbox{\includegraphics[width =7cm]{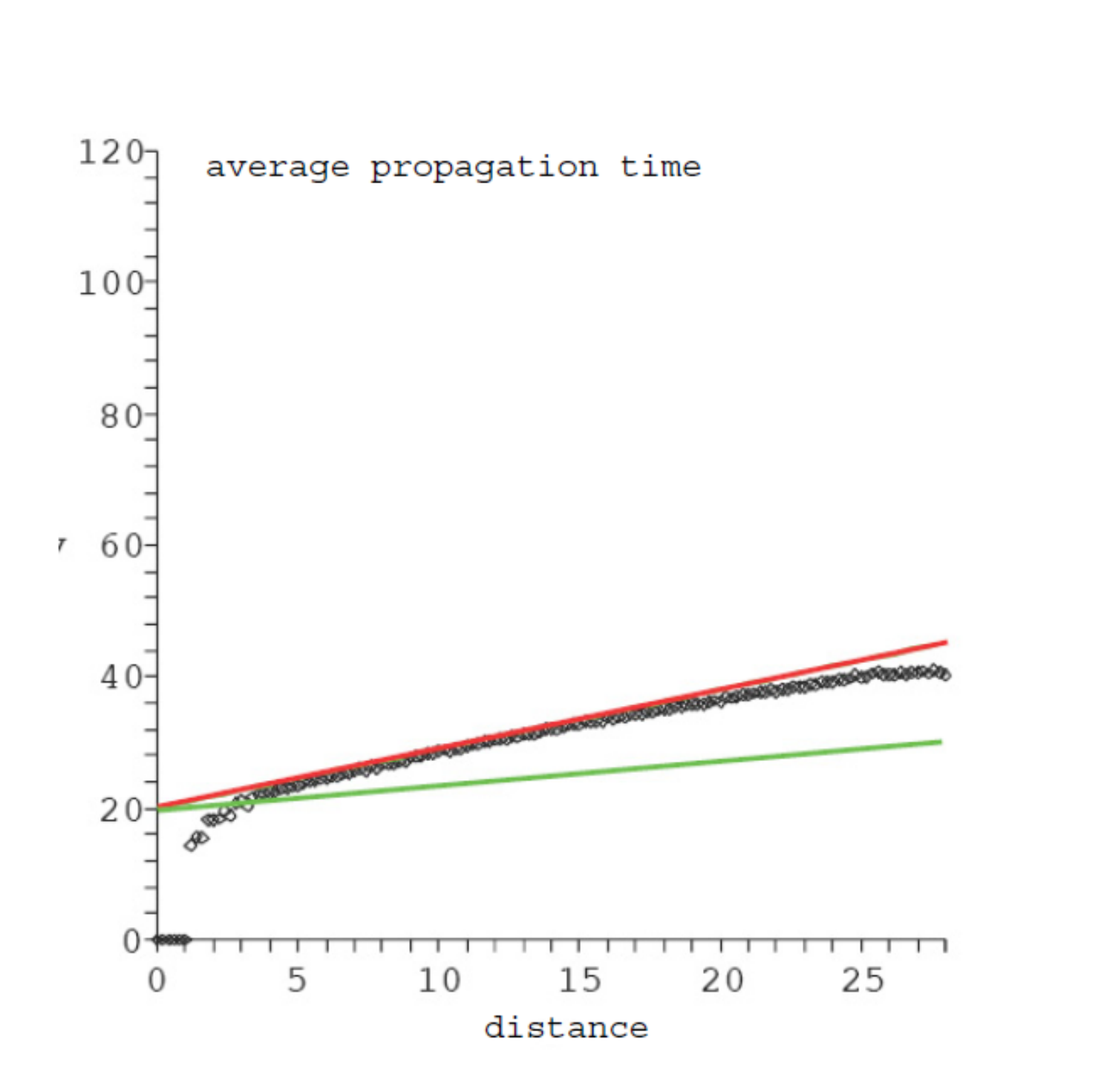}}
\caption{Average propagation time (in seconds) versus distance to source (in meters), compared with theoretical (green - bottom) and heuristic (red - top) slope, for $\tau=0$ and:
$\nu=0.025$ simulated in a $80\times 80$ square (top);
$\nu=0.05$ in a $60\times 60$ square (middle);
$\nu=0.1$ in a $40\times 40$ square (bottom).}
\label{rdwp}
\end{center}
\end{figure}

\begin{figure}[t]
\begin{center}
\fbox{\includegraphics[width =7cm]{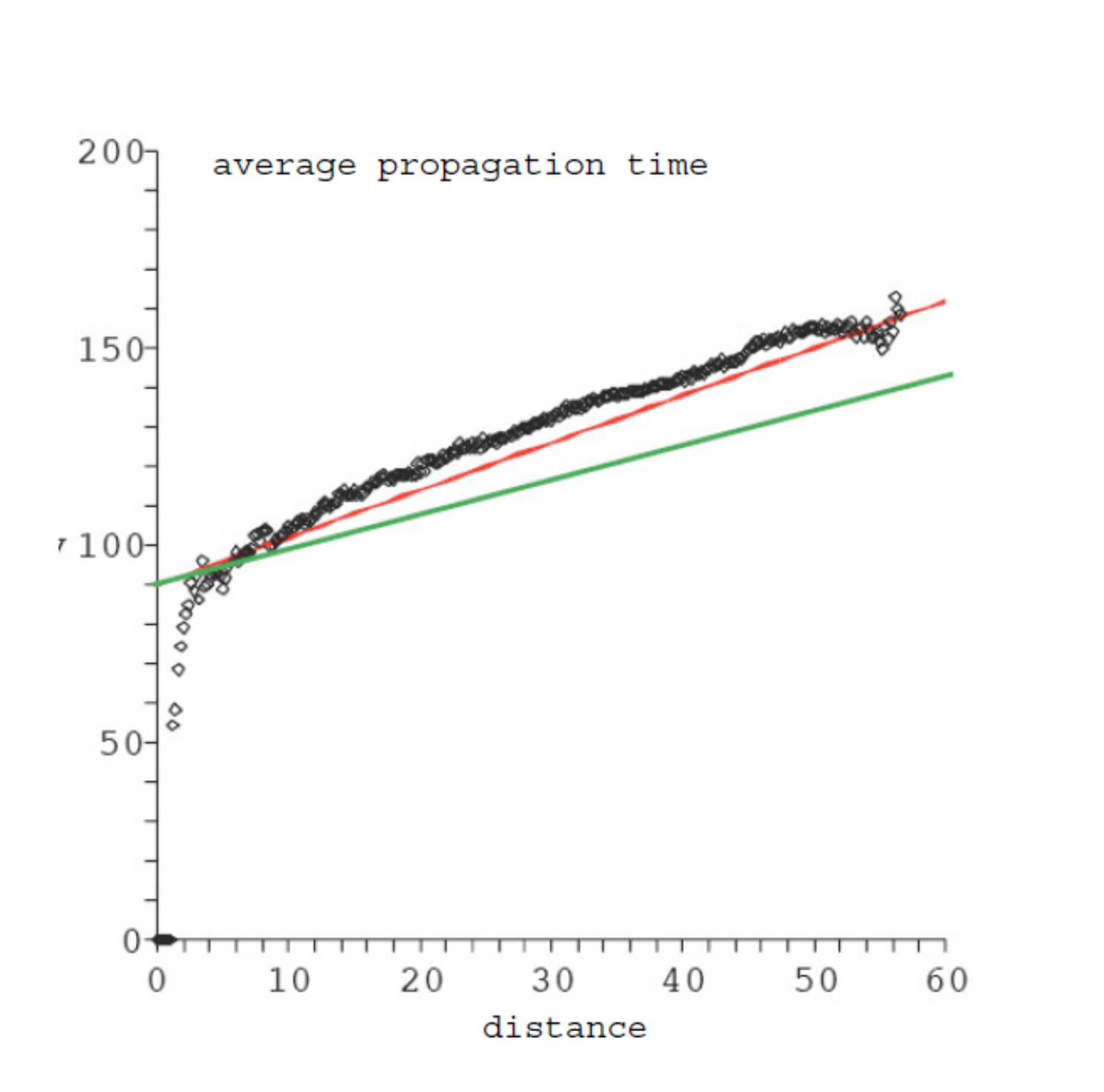}}
\vskip 0.6cm
\fbox{\includegraphics[width =7cm]{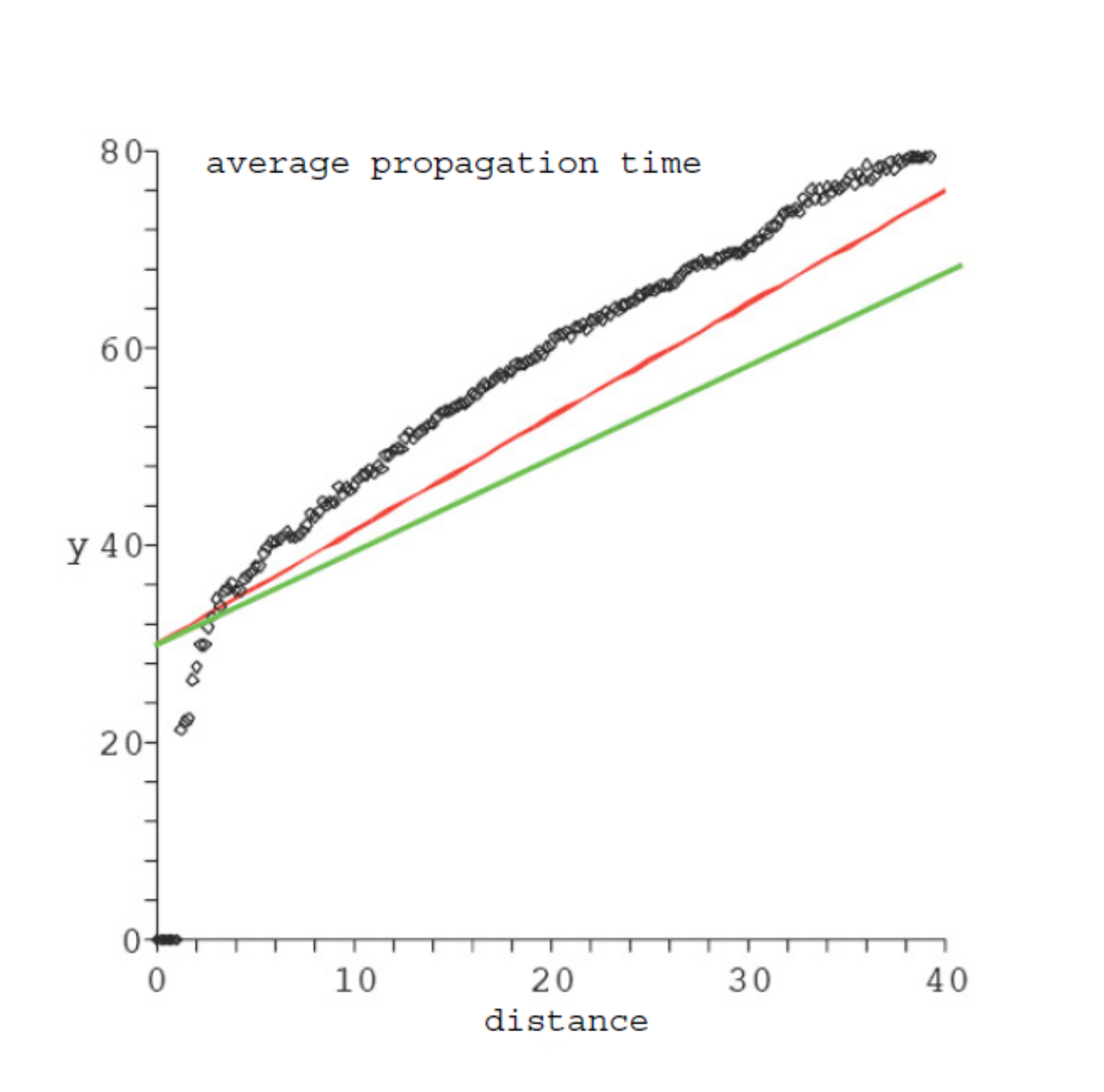}}
\vskip 0.6cm
\fbox{\includegraphics[width =7cm]{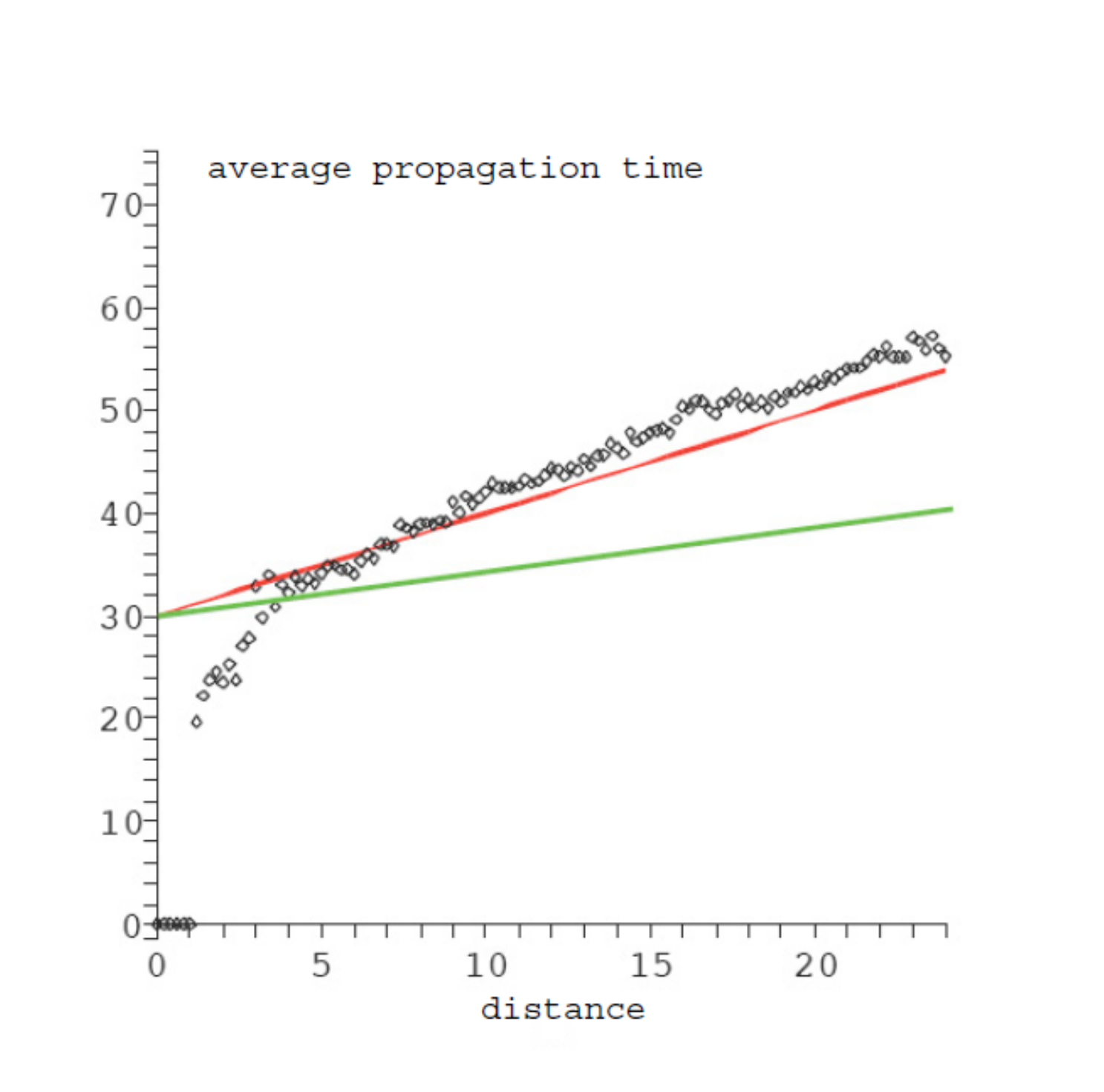}}
\caption{Average propagation time (in seconds) versus distance to source (in meters), compared with theoretical (green - bottom) and heuristic (red - top) slope, for $\tau=0.1$ and:
$\nu=0.025$ simulated in a $80\times 80$ square (top);
$\nu=0.05$ in a $60\times 60$ square (middle);
$\nu=0.1$ in a $40\times 40$ square (bottom).}
\label{rdwk}
\end{center}
\end{figure}
\end{twocolumn}

\end{document}